\documentclass[fleqn,usenatbib]{mnras}

\usepackage[T1]{fontenc}
\usepackage{ae,aecompl}

\usepackage{graphicx}	
\usepackage{amsmath}	
\usepackage{amssymb}	

\newlength{\VSpaceBeforeTabBib}
\newlength{\VSpaceBeforeTabFoot}
\newcommand\tablebib[1]{\VSpaceBeforeTabFoot=1ex%
  \par\vspace{\VSpaceBeforeTabBib}
  \noindent
  \begin{minipage}{\linewidth}
    {\small\bfseries References.}~%
    \small
    \ignorespaces
    #1%
  \end{minipage}%
}
\newcommand\tablefoot[1]{\VSpaceBeforeTabBib=1ex%
  \par\vspace{\VSpaceBeforeTabFoot}
  \noindent
  \begin{minipage}{\linewidth}
    {\small\bfseries Notes.}~%
    \small
    \ignorespaces
    #1%
  \end{minipage}%
}

\title[A temperate 1.5 Earth-mass planet in GJ\,1061]{Red Dots: A temperate 1.5 Earth-mass planet in a compact multi-terrestrial planet 
system around GJ\,1061\thanks{This research has made use of the services of the ESO Science Archive Facility. Based on observations collected at the European Southern Observatory under ESO programmes 072.C-0488(E), 183.C-0437(A), 0101.C-0516(A), and 198.C-0838(A).}}

\author[S. Dreizler et al.]{
S. Dreizler,$^{1}$\thanks{E-mail: dreizler@astro.physik.uni-goettingen.de}
     S.\,V.\,Jeffers$^{1}$,
     E.~Rodr\'iguez$^{2}$,
     M.~Zechmeister$^{1}$,
     J.\,R.\,Barnes$^{3}$,
     \newauthor C.\,A.\, Haswell$^{3}$,
     G.~Coleman$^{4}$,
     S.~Lalitha$^{1}$,
     D.~Hidalgo Soto$^{5}$,
     J.\,B.\,P.\,Strachan$^{6}$,
     \newauthor F.-J.\,Hambsch$^{7}$,
     M.\,J.\, L\'opez-Gonz\'alez $^{2}$,
     N.~Morales $^{2}$,
     C.~Rodr\'iguez~L\'opez$^{2}$,
     \newauthor Z.~M.~Berdi\~nas$^{8}$,
     I.~Ribas$^{9,10}$,
     E.~Pall\'e$^{5,11}$,
     A.~Reiners$^{1}$,
    and G.\,Anglada-Escud\'e$^{6,2}$,
\\
   $^{1}$Institut f\"ur Astrophysik, Georg-August-Universit\"at, Friedrich-Hund-Platz 1, 37077 G\"ottingen, Germany\\
    $^{2}$Instituto de Astrof\'isica de Andaluc\'ia (IAA-CSIC), Glorieta de la Astronom\'ia s/n, 18008 Granada, Spain\\
    $^{3}$School of Physical Sciences, The Open University, Walton Hall, Milton Keynes, MK7 6AA, UK\\
    $^{4}$Universit\"at Bern, Physikalisches Institut, Universit\"at Bern, Gesellschaftsstr. 6, 3012 Bern, Switzerland\\
    $^{5}$Instituto de Astrof\'isica de Canarias (IAC), V\'ia L\'actea s/n, 38205 La Laguna, Tenerife, Spain\\
    $^{6}$School of Physics and Astronomy, Queen Mary, University of London, 327 Mile End Road, London, E1 4NS\\
    $^{7}$Vereniging Voor Sterrenkunde, Brugge, Belgium\\
    $^{8}$Departamento de Astronom\'ia, Universidad de Chile, Camino El Observatorio 1515, Las Condes, Santiago, Chile\\
    $^{9}$Institut de Ci\`encies de l'Espai (ICE, CSIC), Campus UAB, C/Can Magrans s/n, 08193 Bellaterra, Spain\\
    $^{10}$Institut d'Estudis Espacials de Catalunya (IEEC), 08034 Barcelona, Spain\\
    $^{11}$Departamento de Astrof\'isica, Universidad de La Laguna (ULL), 38206 La Laguna, Tenerife, Spain\\
}

\date{Accepted XXX. Received YYY; in original form ZZZ}

\pubyear{2019}

\begin{document}
\label{firstpage}
\pagerange{\pageref{firstpage}--\pageref{lastpage}}
\maketitle

\begin{abstract}
Small low-mass stars are favourable targets for the detection of rocky habitable planets. In particular, planetary systems in the solar neighbourhood are interesting and suitable for precise characterisation.
The Red Dots campaigns seek to discover rocky planets orbiting nearby low-mass stars. The 2018 campaign targeted GJ\,1061, which is the 20$^{\mathrm{th}}$ nearest star to the Sun.
For three consecutive months we obtained nightly, high-precision radial velocity measurements with the HARPS spectrograph. We analysed these data together with archival HARPS data.
We report the detection of three planet candidates with periods of $3.204\pm 0.001$, $6.689\pm 0.005$ and $13.03\pm 0.03$ days, which is close to 1:2:4 period commensurability. After several considerations related to the properties of the noise and sampling, we conclude that a 4$^{\mathrm{th}}$ signal is most likely explained by stellar rotation, although it may be due to a planet. The proposed three-planet system (and the potential four-planet solution) is long-term dynamically stable. Planet-planet gravitational interactions are below our current detection threshold.
The minimum masses of the three planets range from 1.4$\pm 0.2$ to 1.8$\pm 0.3$\,${\rm M_\oplus}$. Planet d, with $m\sin{i} = 1.68 \pm 0.25 {\rm M_\oplus }$, receives a similar amount of energy as Earth receives from the Sun. Consequently it lies within the liquid-water habitable zone of the star and has a similar equilibrium temperature to Earth. GJ\,1061 has very similar properties to Proxima Centauri but activity indices point to lower levels of stellar activity.
\end{abstract}




\section{Introduction}

The Red Dots collaboration\footnote{\url{https://reddots.space/goals/}} is an effort to detect exoplanets orbiting our nearest stellar neighbours, i.e., stars within  5~pc, by concentrating the observational efforts on one star at a time.  
The Red Dots observing cadence ensures the good sampling of signals only partially constrained from previous campaigns. This strategy was decisive in the discovery of  Proxima\,b, orbiting our closest stellar neighbour Proxima Centauri \citep{Anglada2016Natur.536..437A}; it contributed significantly to the discovery of Barnard's Star~b \citep{Ribas2018Natur.563..365R}, and revealed the paucity of terrestrial planet candidates in Barnard's Star. 

Within 5pc, there are 15 exoplanet systems known, mostly around M-stars. Three of them are high multiplicity systems: YZ~Ceti \citep{Astudillo2017A&A...605L..11A}, Wolf~1061 \citep{Wright2016ApJ...817L..20W,Astudillo2017A&A...602A..88A}, and GJ~876 \citep{Rivera2010ApJ...719..890R}.
For planetary systems around low-mass stars similar to  our target GJ~1061 (a late type M-star with low mass of $M=0.12$\,M$_\odot$), state-of-the-art planet formation simulations show that planets form outside the water ice-line. They migrate into warm orbits close to the central star when they attain masses of ~0.5--1\,${\rm M_\oplus}$ \citep{Coleman2017MNRAS.467..996C,Alibert2017A&A...598L...5A}. Typically, these planets migrate in resonant chains and continue to accrete solid material. This means that multi-planet systems of hot and warm Earth-mass and super-Earth planets is a common outcome, with the majority of these systems remaining in or close to resonant configurations. Observed planetary systems around stars similar to GJ\,1061, such as YZ~Ceti, Proxima, TRAPPIST-1 \citep{Gillon2017Natur.542..456G,Grimm2018A&A...613A..68G},  as well as Teegarden's star \citep{2019A&A...627A..49Z} closely match the simulations from recent work which uses both pebble and planetesimal accretion scenarios to form the  planetary systems \citep{2019arXiv190804166C}.

While other programmes such as HIRES/Keck \citep{2017AJ....153..208B} or the HARPS survey 
\citep[e.g. ][]{Astudillo2017A&A...605L..11A} have detected numerous planets by working on large samples of stars, our strategy is different. We perform uniform, nightly cadence spectroscopic and photometric monitoring of a very small number of targets (up to three per season) to cover the periods of temperate orbits several times in each season (up to three months). This strategy has proven optimal \citep{Anglada2016Natur.536..437A} for detection of low-mass planets in hot to temperate orbits around nearby red-dwarfs, and for identifying and mitigating intrinsic stellar variability and spurious correlations  as also demonstrated by \citet{2017A&A...599A.126D}.

 In this paper we first introduce GJ\,1061 with its basic stellar parameters (Sect.\,2) followed by a presentation of the spectroscopic and photometric data (Sect.\,3). In the analysis (Sect.\,4), we first describe the signal detection in the radial velocity data and the detailed analysis of the signals (4.1). In order to assess their origin being due to planetary orbits or stellar activity we then analyse the spectroscopic activity indicies and the photometry (4.2). The results are finally discussed in Sect.\,5.

\section{GJ~1061}
\label{sec:gj1061}

\subsection{General properties of GJ 1061} 

The stellar parameters, including activity indices, are summarised in Table~\ref{tab:stellar-params}. At a distance of $\sim$3.67\,pc \citep{GAIA2018A&A...616A...1G}, GJ\,1061 is the 20$^{\mathrm{ th}}$ nearest star to the Sun\footnote{\url{http://www.recons.org/}} and has
an effective temperature $T_{\rm eff}$ slightly below 3000\,K \citep{Gaidos2014MNRAS.443.2561G,2018AJ....156..102S}. 
Two different $T_{\rm eff}$ values are reported by \citet{2018AJ....156..102S} and \citet{Gaidos2014MNRAS.443.2561G}. While both are consistent, the stellar radius, luminosity, and mass differ between both studies. Since this affects the derived planetary parameters we explicitly indicate in Sect.\,\ref{sec:Analysis} and \ref{sec:Discussion} which parameters were used.

\begin{table}
   \caption{\label{tab:stellar-params}Stellar parameters for GJ\,1061.}
   \centering
   \begin{tabular}{lcc}
      \hline
      \hline 
      \noalign{\smallskip}
      Parameter & Value & Ref. \\
      \hline 
      \noalign{\smallskip}
      Alias name & L 372-58  & \\
      $\alpha$ & 03 35 59.700 & {\em Gaia}\\
      $\delta$  & $-$44 30 45.725 & {\em Gaia}\\
      $\mu_\alpha \cos{\delta}$ [mas/yr]& $745.286\pm 0.118$ & {\em Gaia}\\
      $\mu_\delta$ [mas/yr]& $ $-$373.673 \pm 0.138$ & {\em Gaia}\\
      $\pi$ [mas] & $272.2446 \pm 0.0661$ & {\em Gaia}\\
      $V$ [mag] & $13.06 \pm 0.07 $ & G14\\
      $J$ [mag] & $7.523 \pm 0.02 $ & 2MASS \\
      \noalign{\smallskip}
      Sp. type & M5.5V & H02\\
      $T_{\rm eff}$ [K] & $2953\pm98$, $2999\pm41$ & S18, G14\\
      $L$ [10$^{-3}L_{\odot}$] & $1.7 \pm 0.1$, $3 $ & S18, G14\\
      $R$ [$R_{\odot}$] & $0.156 \pm 0.005$,  $0.19$ & S18, G14\\
      $M$ [$M_{\odot}$] & $0.12 \pm 0.01$, $0.14 $ & S18, G14\\
      {[Fe/H]} [dex] & $-0.08 \pm 0.08$ & N14\\
      \noalign{\smallskip}
      $\gamma$ [km/s] & 1.06 $\pm$0.01 & this work \\
      $v\sin i$ [km/s] & $<2.5$ & this work \\
      $\log L_{\mathrm{H}\alpha}/L_\mathrm{bol}^{\rm *}$ & $<$4.88 (inactive) & this work \\
      $\log L_\mathrm{X}/10^{-7}$\,W &  26.07  & S04 \\
      Age [Gyr]  &    $>7.0\pm0.5$ & W08\\
      \noalign{\smallskip}
      \hline 
   \end{tabular}
\tablefoot{$^{\rm *}$: see \citet{2018A&A...614A..76J} for a conversion from equivalent width to the luminosity ratio as well as for the threshold for inactive stars.}
    \tablebib{
      {\em Gaia}{}: \citet{GAIA2018A&A...616A...1G};
      G14: \citet{Gaidos2014MNRAS.443.2561G};
      2MASS: \citet{Skrutskie2006AJ....131.1163S};
      H02: \citet{Henry2002AJ....123.2002H};
      S18: \citet{2018AJ....156..102S};
      N14: \cite{Neves2014A&A...568A.121N};
      R74: \citet{Rodgers1974PASP...86..742R}; 
      B14: \citet{Barnes2014MNRAS.439..3094B};
      S04: \cite{Schmitt2004A&A...417..651S};
      W08: \cite{West2008AJ....135..785W}.
   }
\end{table}

\subsection{Activity, age, and rotation period estimates}
\label{sec:AAR}

Mid-M dwarf stars generally show significant magnetic activity
\citep[e.g.,][and references therein]{West2008AJ....135..785W, Reiners2012AJ....143...93R, Barnes2014MNRAS.439..3094B, Jeffers2018A&A...614A..76J}. GJ\,1061 is relatively magnetically inactive 
with  little or no signs of H$\alpha$ emission  \citep{Barnes2014MNRAS.439..3094B} and a so far undetermined rotation rate.  \citet{Schmitt2004A&A...417..651S} reported a very low detection of X-ray emission at the level of $\log{L_{\rm X}} \sim 26.07$.  Given its very low activity GJ\,1061 has been used by \cite{Barnes2017MNRAS.471..811B} as a slowly rotating inactive reference for Doppler imaging studies. 

 The chromospheric Ca {\sc ii} index of GJ\,1061 is one of the lowest ($\log{ R'_{\rm HK}}=-$5.32) among  several thousand stars investigated by \citet[][see Fig.~3]{BoroSakia2018A&A...616A.108B}.  This is in agreement with the value $\log {R'_{\rm HK}}=-$5.754 obtained by \citet{Astudillo2017A&A...600A..13A}.  These $\log{R'_{\rm HK}}$ values suggest a rotation period ranging from 50 to 130 days according to the relations by \citet{2015MNRAS.452.2745S}  and \citet{Astudillo2017A&A...600A..13A} and are consistent with the inferred rotation period in the range of 50--200\,d from the X-ray activity-rotation relation by \citet{2003A&A...397..147P}, \citet{Wright2011ApJ...743...48W}, and \citet{Reiners2014ApJ...794..144R}.  These ranges of potential rotation period are consistent with the absence of H$\alpha$ emission \citep{Newton2017ApJ...834...85N} and is consisten with $v \sin i<5$ km\,s$^{-1}$ reported by   
\citet{Barnes2014MNRAS.439..3094B} who as well reported $-6.74 \leq {\rm log}{L_{\rm H\alpha}/L_{\rm bol}} \leq 6.12$ and very low amplitude activity.
From the lack of H$\alpha$ emission we can infer that the star is older than the so-called activity lifetime of an M5 or M6 dwarf, which is 7.0$\pm$0.5\,Gyr \citep{West2008AJ....135..785W}. Neither circumstellar material nor stellar multiplicity has been detected for GJ\,1061 \citep{Avenhaus2012A&A...548A.105A,Rodriguez2015MNRAS.449.3160R} using the WISE and HERSCHEL satellite. Thus, though the star's bulk properties are very similar to those of Proxima Centauri \citep{Anglada2016Natur.536..437A}, GJ\,1061 seems to be less active.
These findings are further supported by the pilot precision radial velocity survey of a sample of 15 M5V - M9V stars conducted over a six night span by \citet{Barnes2014MNRAS.439..3094B}; GJ 1061 showed the lowest variability with r.m.s. variability $< 5$ ms$^{-1}$.

 

\section{Observations}
\label{sec:Obs}

\subsection{HARPS Red Dots}

Observations during 54 nights were made using the HARPS spectrograph at the ESO 3.6\,m at La Silla \citep{Mayor2003Msngr.114...20M} from July to September 2018 with exposure times of 1800\,s and an average signal-to-noise ratio of $\sim20$ at 650\,nm.
We also analyse public archival data ({\em HARPS public}\footnote{ESO programme 198.C-0838(A), PI Bonfils}) treating the seven pre-fiber upgrade and 107 post-fiber upgrade data as separate sets \citep{LoCurto2015Msngr.162....9L}. All reduced spectra were processed with both TERRA \citep{Anglada2012ApJS..200...15A} and SERVAL \citep{Zechmeister2018A&A...609A..12Z} codes, yielding radial velocities (RVs), pseudo equivalent widths (pEW) of H$\alpha$ and Na {\sc i} D, and differential line widths (dLW, see Sect.\,3.2 of \citeauthor{Zechmeister2018A&A...609A..12Z}). 
We use nightly averages in cases of multiple exposures per night resulting in 98 post-upgrade measurements.

\subsection{\label{sec:photo_inst}Photometry}

\begin{table}
    \caption{\label{tab:phot-obslog}Properties of the photometric data 
    sets$^{\rm a}$}
    \centering
    \begin{tabular}{lcrrrr}
        \hline
        \hline
        Data set     &   Season  & $\Delta T$ &$N_{\rm obs}$ &  $N_n$   &    rms~~~\\
                     &           &    [d]     &           &          &      [mmag] \\
        \hline
        ASH2 V       &     2018  &     97     &    426    &   32     &      6.8      \\
        ASH2 R       &     2018  &     97     &    426    &   32     &      8.6      \\
        MONET-S      &     2018  &     96     &   4320    &   43     &     21.2      \\
        AAVSO        &     2018  &    123     &   6987    &  103     &     11.8     \\
        TESS         &     2018  &     53     &  27602    &   44     &      1.5     \\
        MEarth T11   &     2017  &     92     &   1031    &   31     &     10.1      \\
        MEarth T13   &   2016/17 &    243     &  10607    &  171     &      8.8     \\
        ASAS-SN      &   2014/18 &   1558     &   1054    &  407     &     60.5     \\
        \hline
    \end{tabular}
    \tablefoot{
        $^{\rm a}$ Data set identifier, season, time span, number of individual observations, number of nights and rms as average uncertainty over all nights in each data set.
    }
\end{table}

Contemporaneous photometry from ASH2\footnote{Astrograph for South Hemisphere II}, MONET-South\footnote{MOnitoring NEtwork of Telescopes}, and AAVSO\footnote{American Association of Variable Star Observers} was obtained to determine the stellar rotation period of GJ\,1061. 
Moreover, the TESS mission \citep{Ricker2015JATIS...1a4003R} observed this target in continuous mode for $\sim$50 consecutive days during our Red Dots campaign. We also used archival photometry from MEarth \citep{Berta2012AJ....144..145B} and ASAS-SN \citep{Kochanek2017PASP..129j4502K}. These data sets are summarised in Table~\ref{tab:phot-obslog}.  The corresponding observing facilities are described in Appendix\,\ref{app:photofacilities} and Table~\ref{tab:phot-facilities}, the photometric data are listed in Table\,\ref{tab:GB_phot}. 

All new ground-based CCD measurements were obtained by the method of synthetic aperture photometry. Each CCD frame was corrected in a standard way for bias and/or dark, and flat field by instrument specific pipelines. From a number of nearby and relatively bright stars within the frames, the best sets were selected as reference stars.

\section{Data analysis}
\label{sec:Analysis}

\subsection{Signal detection in the radial velocities, and model parameters\label{sec:RVanalysis}}

\begin{figure}
   \centering
   \includegraphics[width=0.52\textwidth]{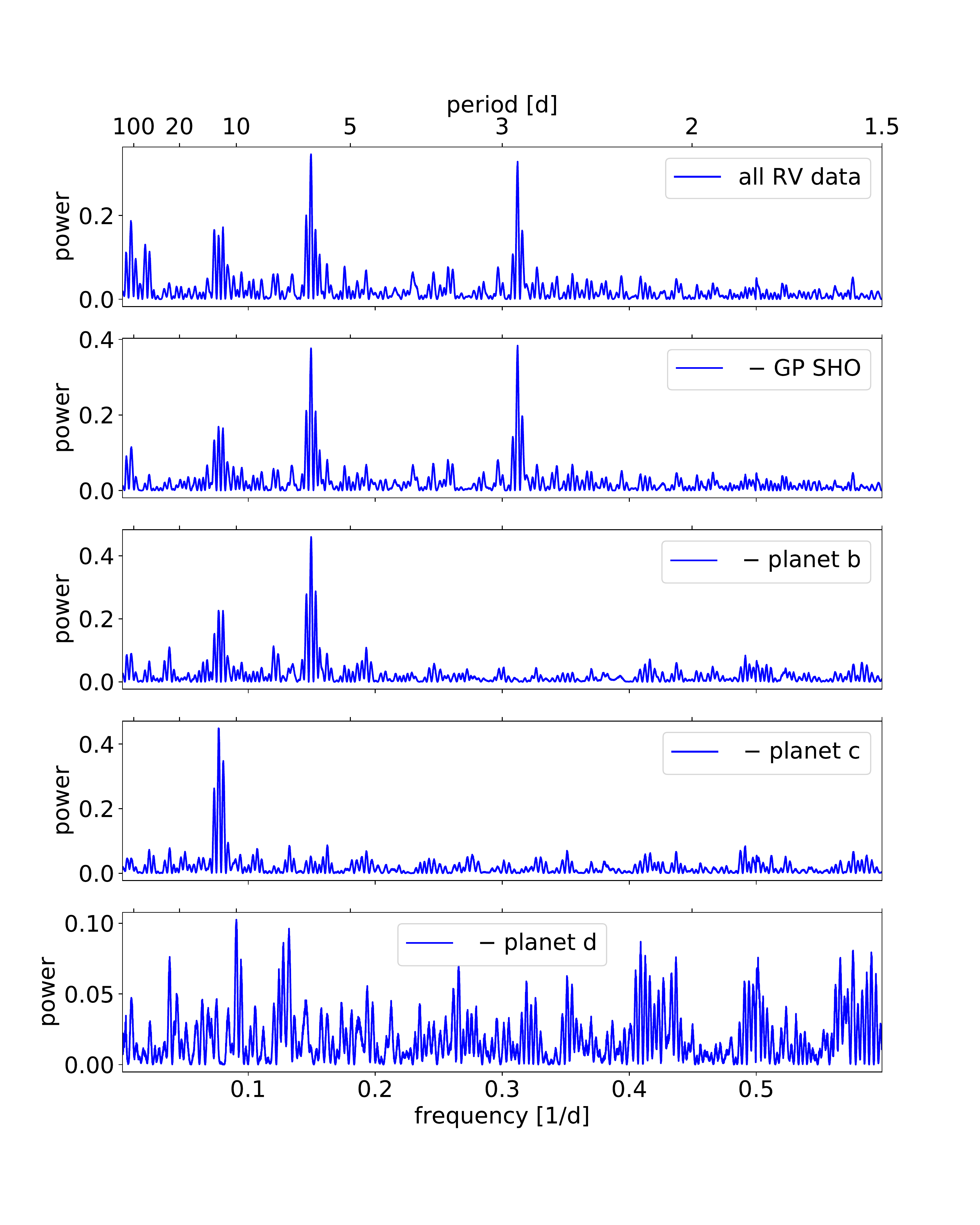}
   \vspace{-12mm}
   \caption{\label{fig:AllRV_prewhite}{\em Top:} Periodograms of HARPS pre-, post-upgrade data with consecutively removed signals.
   The periodograms of the individual signals are shown in the Appendix in Fig.\,\,\ref{fig:AllRV_signalsapp}. }
\end{figure}

The RV analysis incorporates all measurements listed in Tables\,\ref{tab:RV Hpre} and \ref{tab:RV Hpost}, except for one outlier\footnote{The excluded RV point is a 5.5\,$\sigma$ outlier from JD=2458376.8825. The spectrum shows no anomaly except the +20.45\,m/s RV deviation. It is listed in Table~\ref{tab:RV PRD} for completeness.}. 
The first step of the analysis is to detect the most likely periodic signals in the data and assess their significance using likelihood periodograms \citep{2009MNRAS.393..969B}. We use an iterative sequential procedure consisting of: i) computing a likelihood periodogram to identify the period of the (next) most probable signal, ii) including the signal in the model and repeating the process until no significant signals remain.  Using our favoured model described below, this procedure is illustrated in Fig.~\ref{fig:AllRV_prewhite}, showing four detected signals. Sometimes a new period is significant, but ambiguous due to aliasing. To ensure a secure and unique period determination, we additionally require that the maximum likelihood solution (highest peak) must have a $\Delta$ ln L $>$ 5 compared to the second highest peak. In a Bayesian sense, this is equivalent to requiring that the model probability (assuming uniform priors) of the best solution must be $e^5 \sim 150$ times more probable than any other solution with an alternative period and the same number of parameters. In addition to Keplerian signals, the model also includes a zero-point offset and a white noise jitter parameter for each set, a global linear trend, and a model for the correlated noise (see below).
We also used the {\em Systemic} \citep{Meschiari2009PASP..121.1016M} interface to visually verify that the solution was indeed unique.

\begin{figure}
    \centering
    \includegraphics[width=0.52\textwidth]{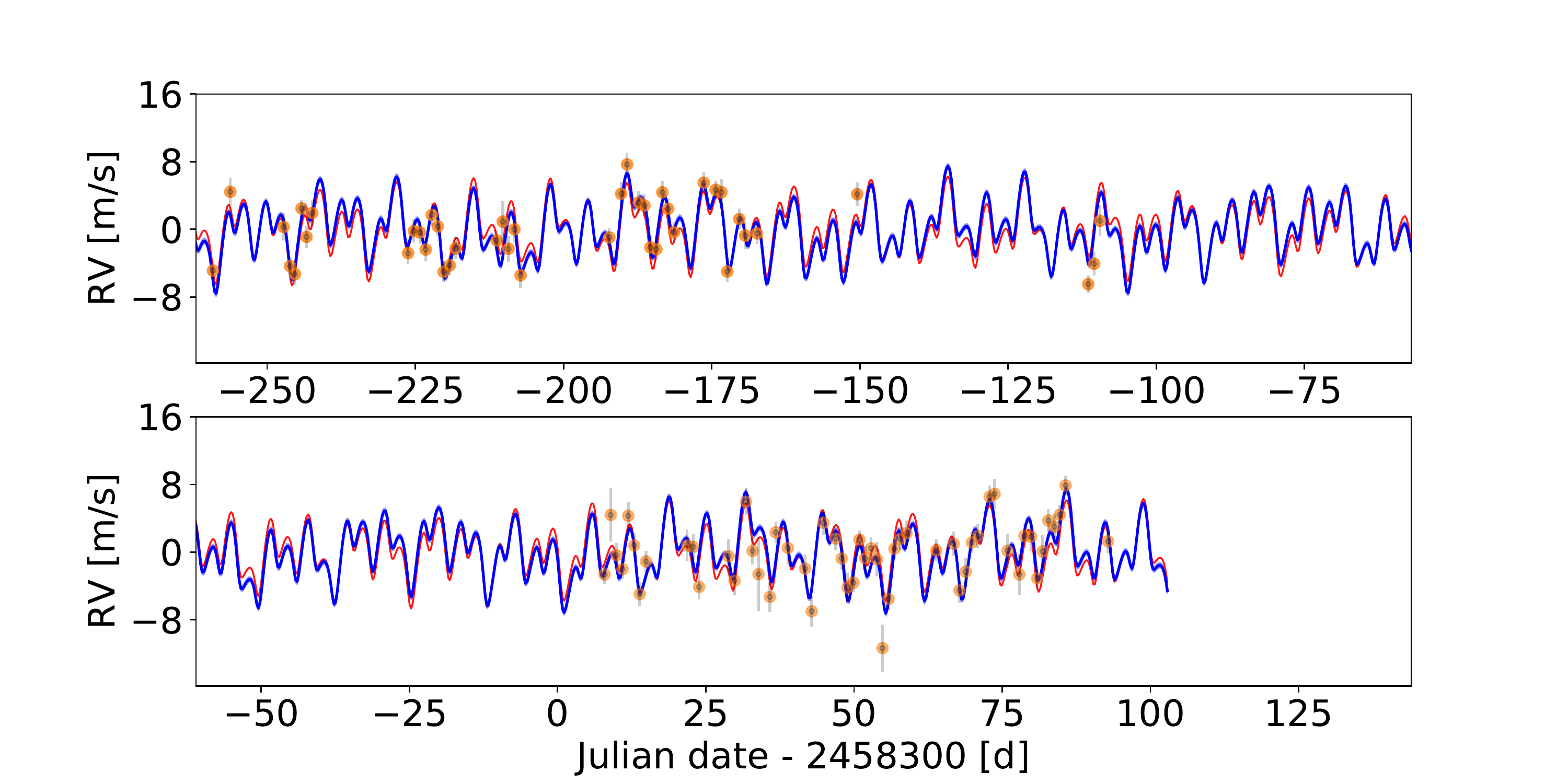}
    \vspace{-5mm}
    \caption{HARPS public and Red Dots data (from BJD 2458300 on) measurements with our favoured model with 3-planet model (red). The blue line includes the modeling of the correleated noise using the {\em SHO} kernel from {\em celerite}.}
    \label{fig:fit}
\end{figure}
 In the second step, i.e. the detailed analysis, we deployed a python script using {\em celerite} \citep{2017AJ....154..220F} and {\em emcee} \citep{Foreman-Mackey2013PASP..125..306F} to characterize the credibility intervals for all the parameters in the model.
 For an increasing number of planets ($N=0,1,2,3,4$) the parameters for each Keplerian model are the orbital period $P$, the radial velocity semi-amplitude $K$, the eccentricity $e$, the periastron longitude $\omega$ as well as the periastron passage time $t_{\rm peri}$\footnote{The eccentricities of the first three Keplerian signals are small. Allowing eccentric Keplerian solutions for them does not significantly improve upon circular orbit fits.}. The 4$^{\mathrm{th}}$ signal was alternatively modelled using the stochastically driven damped harmonic oscillator kernel ({\em SHO}) provided by {\em celerite} to account for correlations between data points using a Gaussian Process (GP) framework (see Appendix\,\ref{App:models} for more details). Model parameters for this kernel are an amplitude $S_0$, a resonance frequency $\omega_0$ (we convert it in a period $P_0$), and the quality factor $Q$ of the oscillator converted to a damping time scale, $\tau_{\rm d}$. Like in the first step, the model also includes a zero-point offset and a white noise jitter parameter\footnote{The jitter parameter for the HARPS pre-upgrade data result in an unrealistic high value of about 10\,m/s which is likely due to the fact that this data set only contains seven data points scattered over about 10 years. We therefore fixed this jitter parameter to 1\,m/s. This has no impact on the overall results.} for each instrument, and a global linear trend. 

The first and second signals (planets~b and c) are unique with no severe aliasing ambiguities. However, for the third signal (planet~d), the period could be either 13.0\,d or 12.4\,d given that the $\Delta \ln L$ between the two solutions is only about $2.5$. In frequency space, the difference between $1/13.0\,{\rm d}^{-1}$ and $1/12.4\,{\rm d}^{-1}$ is $1/270$, which corresponds to the highest secondary lobe of the window function for this time series, thus the confusion is probably caused by aliasing. The bulk properties of the planet candidates (period and mean longitude) do not change dramatically between solutions but the amplitude and therefore mass is affected, although within error bars. Further observations can to pin-down the precise period and orbital elements. Both solutions are listed in Table~\ref{tab:orbit}. 

A similar, but more problematic, situation occurs with the 4$^{\mathrm{th}}$ signal. From the periodograms we cannot robustly distinguish between P=53\,d and P=130\,d because these two solutions only differ by a $\Delta \ln L $ of $\sim$3; including either in the model removes the other, indicating an aliasing between the two. Treating the 4$^{\mathrm{th}}$ signal as a Keplerian orbit results in a rather high eccentricity of e=0.34. When modelled with the {\em SHO}-kernel, the fit favours a period of P$_0=53$\,d rather than 130\,d and an unconstrained long damping time $\tau_{\rm d}>130\,d$. Since the period of the 4$^{\mathrm{th}}$ signal is overlapping with the probable range for the rotation period, we discuss the interpretation of that signal as due to activity or planetary orbit in Section\,\ref{sect:ActivityAll}.

We also tested whether a $N=4$ model with $P_4=53$\,d including an {\em SHO} kernel would result in a statistically significantly better fit. This is not the case since the log-likelihood increases only by 3.
By comparison of the log-likelihood of the different models ($N=0\ldots4$, {\em SHO})and taking potential activity induced RV signals into consideration (see below) we identified our favoured model as the one with three planets and a {\em SHO} kernel to describe the correlated noise.

 Table~\ref{tab:orbit} summarizes the most probable parameter values, their credibility intervals, and the statistical significance of each signal for our favoured model with three Keplerian signals and the {\em SHO} kernel. The improvement of the fit compared to $N-1$ signals is indicated as $\Delta \ln L$. Statistical significances are given through the False Alarm Probability estimates derived from the improvement in the log-likelihood statistic \citep{2009MNRAS.393..969B}. Fig.\,\,\ref{fig:fit} shows the best fit for all post-upgrade HARPS data. 
 Fig.\,\,\ref{fig:RVphase} shows the final phase-folding with the model specified in Table~\ref{tab:orbit}. The periodograms of the data with these four signals subsequently removed are shown in Fig.\,\ref{fig:AllRV_prewhite}, the periodograms of the corresponding models are shown in Fig.\,\ref{fig:AllRV_signalsapp}. The posterior distribution of the parameters are shown in the Appendix in Figs.\,\ref{fig:MCMC_full} to \ref{fig:MCMC_d}. 

As a final check, we investigated the phase coherence using data from 2017 (HARPS public) and 2018 (HARPS Red Dots) separately. The first three signals are detected in both sub-sets with parameters matching the full analysis, however, larger uncertainties (about factor 10 in period and factor 2 in mass). Neither season is sufficiently long for a clear detection of the fourth signal. This is further illustrated with Fig.\,\,\ref{fig:SBGLS} in Appendix\,\ref{app:periodogram}.
The signals from the three planets are coherent over two seasons, i.e.~these detections are robust. 

\begin{table*}
   \centering
   \caption{\label{tab:orbit} Model and reference statistical parameters from the simultaneous fit of HARPS pre and post upgrade data for the 3-Keplerian signals plus {\em SHO} model. Although we fit full Keplerian orbits, they are indistinguishable from circular ones (Fig.\,\ref{fig:MCMC_b}-\ref{fig:MCMC_d}). The derived values for semi-major axis and planetary mass take the stellar mass uncertainty listed in Table\,\ref{tab:stellar-params} into account. Due to aliasing, we list two solutions for planet~d (solutions with P$_{\mathrm{d}}=12.4$ denoted with *. $\Delta\ln L$ indicates the fit improvement with respect to the model with one less planet.
   False Alarm Probabilities (FAPs) are calculated from the improvement of the log-likelihood statistic following \citet{2009MNRAS.393..969B}.}
   \begin{tabular}{lcccc}
      \hline 
      \hline 
      \noalign{\smallskip}
      Keplerian parameters & Planet b & Planet c & Planet d & Planet d* \\
      \noalign{\smallskip}
      \hline 
      \noalign{\smallskip}
      Detection order & \#1 & \#2 & \#3 & \#3* \\
      $P$ [d]        &$    3.204\pm 0.001 $ & $6.689\pm 0.005$ & $ 13.031^{+0.025}_{-0.032}$& $12.434^{+0.031}_{-0.023}$\\[0.3em]
      $K$ [m/s]      &$      2.43\pm 0.24 $  & $2.48^{+0.28}_{-0.29}$ & $ 1.86\pm 0.25$ &                         \\[0.3em]
      $K$* [m/s]     &$      2.54\pm 0.25 $  & $2.39\pm 0.24$ &                         & $ 1.76 \pm 0.28$ \\[0.3em]
      $e$ [-]$^{\rm a}$ &$    <0.31 $         & $<0.29$ & $<0.53$ & $<0.54$ \\[0.3em]
      $\omega$ [deg] & $145^{+81}_{-65}$ & $88^{+95}_{-85}$ & $157^{+88}_{-71}$ & $278^{+49}_{-214}$\\[0.3em]
      $t_{\rm peri}$ [d]$^{\rm b}$ & $ 0.61^{+0.56}_{-0.72} $ & $ 0.2^{+1.9}_{-1.5} $ & $6.4^{+3.1}_{-2.5} $ & $8.0^{+2.0}_{-2.4}$ \\[0.3em]
      \noalign{\smallskip}
      \hline 
      \noalign{\smallskip}
      \multicolumn{2}{l}{Derived parameters}\\
      \noalign{\smallskip}
      \hline 
      \noalign{\smallskip}
     $a$ [au]                &$0.021 \pm 0.001$ & $ 0.035\pm 0.001$ & $0.054\pm0.001$ & $0.052\pm 0.001$\\
      \noalign{\smallskip}
      $m \sin i$  [$\rm M_\oplus$] & $1.38 ^{+0.16}_{-0.15}$ & $1.75\pm 0.23$ & $1.68^{+0.25}_{-0.24}$ & \\[0.3em]
      $m \sin i$* [$\rm M_\oplus$] & $1.44^{+0.17}_{-0.16}$  & $1.74\pm 0.20$ &                        & $1.57^{+0.27}_{-0.25}$ \\[0.3em]
      $\lambda$ [deg]$^{\rm b,c}$ &$    66.0\pm 5.7 $         & $80.0\pm 7.6 $ & $335.7^{+8.4}_{-8.0}$ & $72.2\pm8.9$ \\[0.3em]
      $t_{\rm c}$ [d]$^{\rm b,d}$ & $3.4\pm 0.1$ & $6.9\pm 0.2$ & $4.2\pm 0.7$ & $13.1\pm 0.7$\\[0.3em]
      $F$ [$\rm S_\oplus$]$^{\rm c}$  & $3.8\pm0.7$ & $1.4\pm 0.2$ & $0.6\pm 0.1$ & $0.6\pm 0.1$ \\[0.3em]
      \noalign{\smallskip}
      \hline 
      \noalign{\smallskip}
      HARPS pre [m/s] &  \multicolumn{4}{c}{$1.0^{\rm d}$} \\
      \noalign{\smallskip}
      HARPS post \& Red Dots [m/s] & \multicolumn{4}{c}{$1.06^{+0.18}_{-0.17}$} \\
      \noalign{\smallskip}
      $\Delta\ln L$ & 16.8 & 19.1 & 23.3 & \\
      FAP [10$^{-5}$] & 4.5 & 1.6 & 0.12 & \\
      \noalign{\smallskip}
      \hline 
   \end{tabular}
   \tablefoot{
      $^{\rm a}$ Upper limit to orbital eccentricities (99\% credibility interval).
      $^{\rm b}$ Reference time is BJD=2458300.
      $^{\rm c}$ Mean longitude. 
      $^{\rm d}$ Time of inferior conjunction.
      $^{\rm e}$ Insolation with stellar parameters adopted from \citet{2018AJ....156..102S}.
      $^{\rm f}$ fixed.
   }
\end{table*}

\subsection{Activity or planets?}
\label{sect:ActivityAll}

After the signal detection, we assess the validity of the interpretation of the signal as being due to planetary orbits or due to stellar activity. We employ spectroscopic activity indices as well as photometry in the following.

\subsubsection{Spectroscopic indices}
\label{sect:Activity}

We measured pseudo-equivalent widths (EW) of the two Na {\sc i} D lines and H$\alpha$ chromospheric lines as obtained with {\em TERRA} (Table\,\ref{tab:EWs}) to search for evidence of periodicities in the stellar activity (Fig.\,\,\,\ref{fig:Halphaperiodogram}). For this investigation, we excluded epochs with EW measurements showing significant deviations from the quiescent state of the chromosphere of the star (occasional small flares, cosmic rays, and background contamination can all cause spurious measurements of EW) through a two-step $3\sigma$ clipping. These excluded measurements are flagged with $0$ in the online version of the data tables. The H$\alpha$ index shows a signal at about 130\,d which we can well model (Fig.\,\,\ref{fig:Halpha}) with a Gaussian Process using the {\em SHO} kernel described above. 

\begin{figure}
   \centering
   \includegraphics[width=0.49\textwidth]{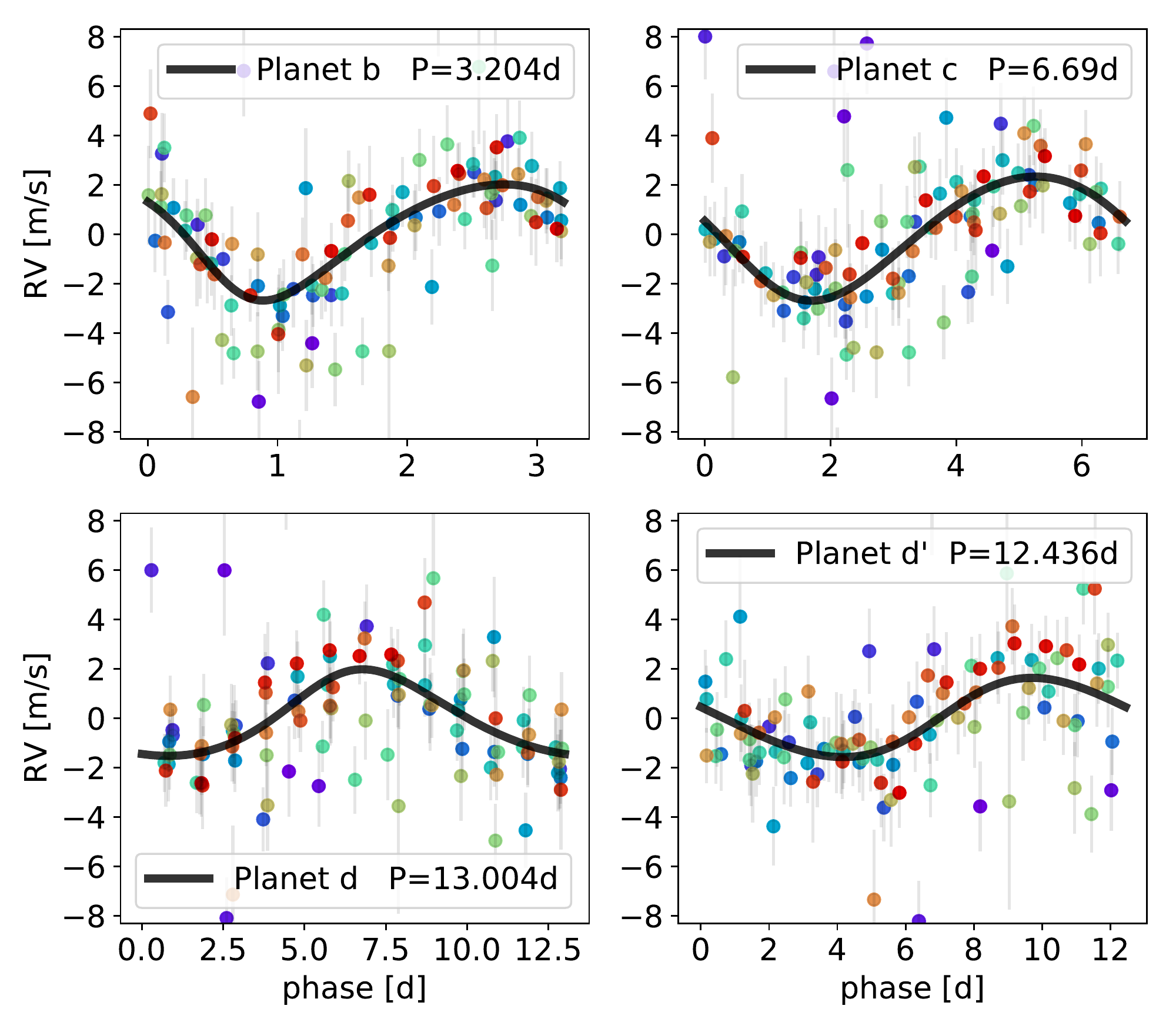}
   \caption{\label{fig:RVphase}Phased HARPS Red Dots data for the 4-signal fit for the 3 planets. Top panels are for planets b and c, and bottom panels are for the two possible solutions for planet d with periods of 13.0 and 12.4\,d. The fourth signal is modeled using the {\em SHO}-kernel. The color indicates the time of observation (chronological order from blue to red) to illustrate the coherence and of the signals along the campaigns.}
\end{figure}

Changes in the width of the average spectral line are known to correlate with spurious Doppler shifts. We also investigated differential change in the line width (or dLW, Table\,\ref{tab:RV Hpost}) from {\em SERVAL} \citep{Zechmeister2018A&A...609A..12Z}. The periodogram of this index (Fig.\,\,\ref{fig:dlwperiodogram}) show two possible signals at 27\,d and 13.5\,d, but at rather low significance (FAP$\sim$1-10\%). The latter is close to, but not exactly at the period of one of the RV signals, so we examined the issue further. We fitted the dLW-data with three models: a Keplerian model accounting for strictly periodic but not necessarily sinusoidal variations; and with two different kernels provided by {\em celerite} to model correlations between data points. The {\em REAL} kernel describes correlation through an exponential excitation or decay at a characteristic time scale, the {\em SHO} kernel is as described above (Sect.\,\ref{sec:RVanalysis}). The fit with a periodic Keplerian-like signal yields 26.07$\pm 0.04$\,d, where the eccentricity also accounts for the signal at 13.5\,d (first harmonic). However, this signal is not coherent over all data, but only for about one observing season of about 100\,d length. The fit of the dLW (Fig.\,\,\ref{fig:dlw}) using both kernels results very short characteristic and damping timescales (of few days), which further supports that the coherence of the signal is short lived. On the other hand, the third RV signal is very coherent throughout the season (see colour coded phase folded plots Fig.\,\ref{fig:RVphase}), so we consider it very unlikely that the dLW and RV signals are related at a significant level. The dLW signal does not match the photometric period either (Sect.\,\ref{sec:Photo}) but it is about half of the 53\,d period present in the RV data. This raises doubt on the planetary origin of the 53\,d signal. 

It should be noted that the dLW may also be due to instrumental or observational effects. Even when the star is very inactive, spurious signals can arise from varying Moon contamination (periodicity close to the detected  $\sim 27$\,d), varying background and therefore spectral line contrast, or small instrumental variations such as small focus changes.

\subsubsection{Photometry analysis}
\label{sec:Photo}

{\em TESS} \citep{2015JATIS...1a4003R} observed GJ\,1061 in Sector 3 and 4 (TIC79611981). With the planetary radii derived in Sect.\,\ref{sec:Discussion} we expect transit depths between about 3 and 40 mmag, depending on the planetary composition. Using the Mikulski Archive for Space Telescopes (MAST)\footnote{\url{https://mast.stsci.edu/}}, we inspected the {\em TESS} data 
for transits, finding no transit signal at any of the RV periods. TESS data has a too short time-span ($\sim$ 60 days) for a detection of the potential long-term rotation modulation in GJ\,1061.

We combined most ground-based time series (omitting ASAS-SN due to large rms and ASH2 V) that we could find on the star, including the new Red Dots photometric data obtained (quasi-)simultaneously to the HARPS observations (Fig.\,\,\ref{fig:photo}). The combined analysis using an {\em SHO} kernel as well as individual offsets and jitter terms shows several strong and clear signals in the range of 60--150\,d (Fig.\,\,\ref{fig:photoperiodogram}, top). We are aware of the fact that a combination of ground-based photometry using different filters as well as comparison stars is prone to artefacts. We therefore also investigate the data sets individually. ASH2 R as well as AAVSO show a long-term periodicity which is however unconstrained due to insufficient duration. In the MEarth T13 data set periodicity at 130$\pm5$\,d is seen most clearly. The 130\,d signal is not coherent over the full photometric data set; this is consistent with an origin in relatively short-lived active regions on the slowly rotating stellar surface. The MEarth T13 data set also shows power in the 5\,d and 2.5\,d range at low amplitude. These, however, disappear in the residuals (Fig.\,\,\ref{fig:photoperiodogram}, bottom).

Using the {\em SHO} kernel, we estimate a damping time $\tau_{\rm d}\sim 20$\,d, i.e. significantly shorter than the underlying oscillator period of $\sim$130\,d further supporting this picture. Alternatively, the {\em REAL} kernel yields $\tau_{\rm d}$ $\sim 44$\,d. Correlation kernels following an over-damped oscillator produce signals in a broad period range as discussed in \citet{Ribas2018Natur.563..365R}\footnote{As a word of caution we like to note that the parameters of the GP models may not have a physical meaning but due to instrumental systematics or due to the combination of inhomogeneous data like in this case.}.

We therefore assessed the chances of the 4$^{\mathrm{th}}$ signal being a false positive induced by correlated noise with the characteristics of those of the star activity traced by the photometry. Since the period of 130\,d being detectable in the  RV, photometric, and  H$_\alpha$ activity index data, we concentrate on the 53\,d signal in the following.
We generated 50\,000 realisations of correlated noise using the {\em REAL} kernel with a $\tau_{\rm d} \sim 44$\,d at the sampling of the RV data. For each of those we obtained the highest peak from a GLS periodogram \citep{Zechmeister2009A&A...496..577Z}. As expected, we find a broad distribution of periods among those peaks with a broad maximum in the range 100 to 400\,d. A signal with a period of between 51 and 57\,d (the full range of periods from the $N=4$ planet fit) is found with 16\,\% chance. For the shorter orbital periods a false-alarm probability is negligible. While a planetary origin for the 53\,d signal cannot be ruled out, we consider stellar activity is the most likely explanation for this 4$^{\mathrm{th}}$ signal.

Summarizing, both, the 53\,d and the 130\,d signal are within the expected range for the stellar rotation period as mentioned in Sect.\,\ref{sec:AAR}. Both have possible counter parts in the activity indices (twice the period of the dLW) or photometry. The common detection of the 130\,d signal in RV, photometric, and activity index data justify to identify the 130\,d period as the stellar rotation period.

\subsection{A stable compact planetary system}

We tested the dynamical properties of the model using {\em mercury6} \citep{Chambers1999MNRAS.304..793C} and {\em Systemic}, and assuming four planets as the dynamically most challenging case. Consistent with expectations based on separations compared to mutual Hill radii, and the period commensurabilities  (close to 1:2:4:16), the system survives at least $10^8$ years. This is, however, requiring small eccentricities for planets~b to d, especially planet~c needs an eccentricity near zero in order to keep the system long-term stable. We found that, when adopting the minimum planet masses, the system is not in a mean-motion resonance since the libration angles rotate. 
Significant exchanges of angular momentum are nevertheless present and manifest as regularly oscillating eccentricities. A dynamical fit results in a very subtle shift in orbital periods within the error bars which, over the duration of observation, causes a slight phase shift without corresponding to a statistically better model. The planetary system remains stable even if viewed almost face-on, at an inclination of $i=10^\circ$.
This dynamical robustness precludes strong constraints on the system's inclination. For randomly oriented systems, geometrical arguments disfavour small values of $i$ and thus masses larger than 3-4 times the reported minimum masses. Probabilistic arguments  based on the Kepler mission's detection statistics and comparison to RV detected samples \citep[e.g.][]{bixel2017ApJ...836L..31B} makes it {\it a priori} very unlikely that these planets have masses larger than 3\,${\rm M_\oplus}$.

\section{Discussion and conclusions}
\label{sec:Discussion}

\begin{figure}
   \centering
   \includegraphics[width=1\linewidth]{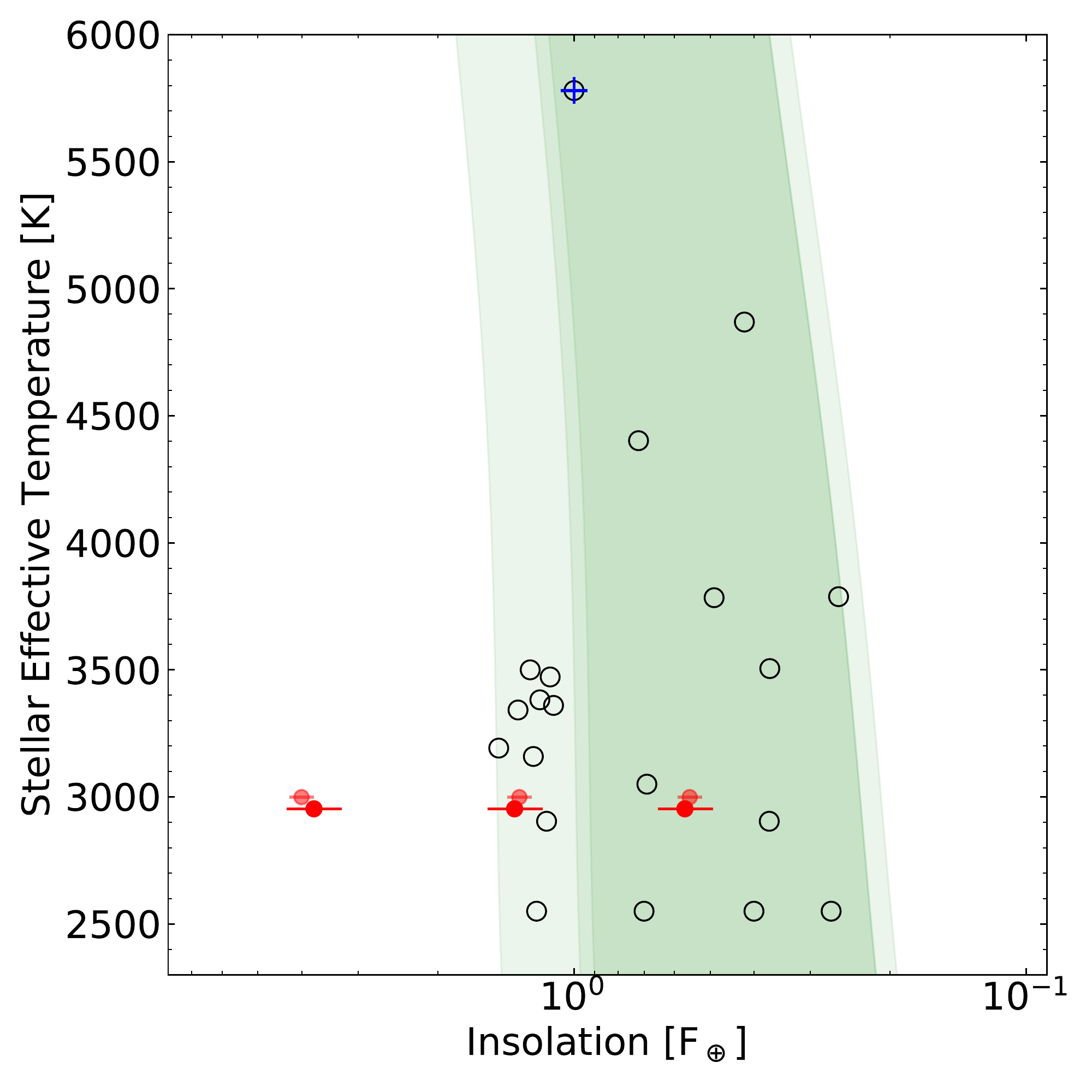}
   \caption{\label{fig:HZ} The optimistic (light green) and conservative (dark green) habitable zone for $1\,{\rm M_\oplus}$ according to \citet{Kopparapu2014ApJ...787L..29K}. GJ\,1061 planets b to d are shown as red dots using the stellar parameters from \citet{2018AJ....156..102S} and as light red dots using values from \citet{Gaidos2014MNRAS.443.2561G}. For comparison we plot selected planets (including Proxima b \citeauthor{Anglada2016Natur.536..437A} \citeyear{Anglada2016Natur.536..437A}, TRAPPIST-1 e, f, and g \citeauthor{Gillon2017Natur.542..456G} \citeyear{Gillon2017Natur.542..456G}, Teegarden's star b and c \citeauthor{2019A&A...627A..49Z} \citeyear{2019A&A...627A..49Z}, and GJ\,357\,d \citeauthor{2019arXiv190412818L} \citeyear{2019arXiv190412818L}) 
   and the Earth (black-blue circled plus) at the top.}
  
\end{figure}

\begin{figure}
   \centering
   \includegraphics[width=1\linewidth]{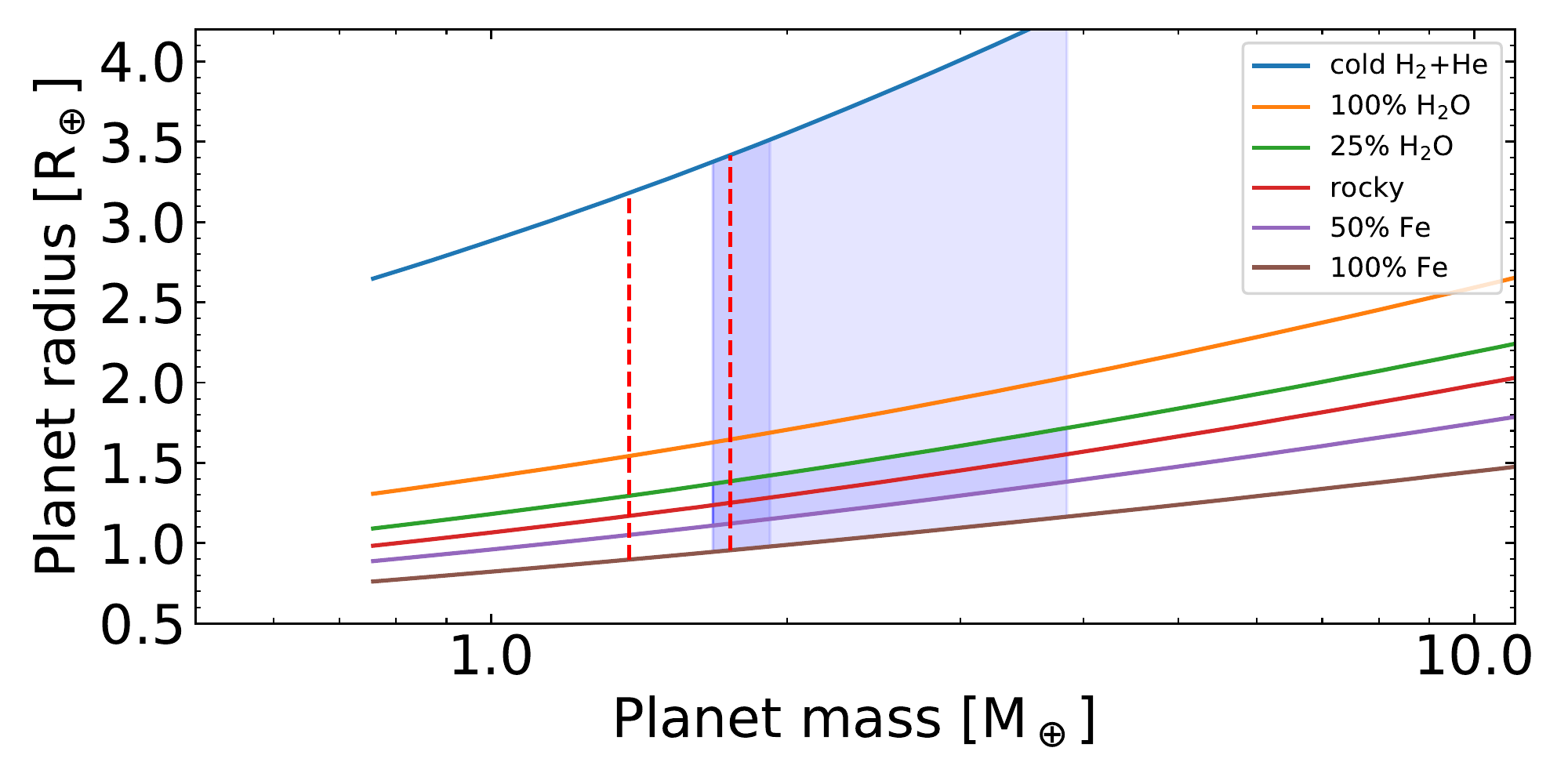}
   \caption{\label{fig:MR} Mass-radius-relation for various compositions \citep{Zeng2016ApJ...819..127Z}. The range of masses up to 90\% cumulative probability of planet~d is shaded in light blue. The region of less extreme compositions between 25\,\% H$_2$O and 50\,\% Fe is additionally highlighted. Stellar parameters are taken from \citep{2018AJ....156..102S}. Additionally shown are the lower masses for the other planets (red dashed lines).}
\end{figure}


We find three RV signals that can  be attributed to planets with minimum masses close to the mass of the Earth matching the expectation from state-of-the-art planet formation simulations \citep{Coleman2017MNRAS.467..996C,Alibert2017A&A...598L...5A}. All three signals are coherent over the observational time span $\sim$1\,year. A fourth signal is also significant, but considering all available data we find that it is mostly likely caused by correlated noise induced by stellar activity and rotation; we find $P_{\rm rot} \sim 130\,$d,  since this signal is detected in the RV, photometric as well as H$_\alpha$ index data. Whether the signal at 53\,d is only due to correlated noise or planetary origin would require longer-term observations with dense coverage over about five months in order to check its coherence as well as amplitude stability. 

Using the stellar parameters  (Table\,\ref{tab:stellar-params}) and  planetary parameters (Table\,\ref{tab:orbit}) we show the three planets in relation to the star's habitable zone (Fig.~\ref{fig:HZ}), assuming the minimum mass as the actual mass. Fig.\,\ref{fig:HZ} shows their insolation, that of  other Earth-like planets, and the limits of the optimistic and conservative habitable zone \citep{Kopparapu2014ApJ...787L..29K}. 
The two sets of stellar parameters result in different insolation values for the planets. We propagate the uncertainties in $T_{\rm eff}$ into the uncertainties in  insolation values. 
With an orbital period of around 13\,d, planet d is the most interesting  regarding potential habitability as it lies well within the classic liquid water habitable zone for both sets of stellar parameters. The range of possible sizes 
for planet~d are shown in Fig.\,\ref{fig:MR}
for various planetary composition ranging from a cold H+He to a pure Fe composition \citep{Zeng2016ApJ...819..127Z}. 


This study illustrates the efficiency of the Red Dots observing strategy. Our homogeneously sampled data over one season has been essential to the robust detection of the three short period planets. Additional planets in longer-period orbits may also be present in the system. The non-zero linear trend of 1.8$\pm 0.4$\,m\,s$^{-1}$\,yr$^{-1}$(Fig.\,\ref{fig:MCMC_jitter}) is indicative of a possible long-period companion, but  a much longer observational baseline is needed to explore this. The sequence of period ratios close to 2 for the inner three planets may hint at an additional undiscovered planet exterior to planet d. Sub-Earth mass planets in periods longer than 20\,d are below the current detection threshold, but could be revealed with further high precision RV measurements.

\section*{Acknowledgements}
S.D., S.J., A.R., L.S., and M.Z. acknowledge support from the Deutsche Forschungsgemeinschaft under Research Unit FOR2544 ``Blue Planets around Red Stars'', project no. DR 281/32-1.
S.J. acknowledges project JE 701/3-1 and DFG priority program SPP 1992 ``Exploring the Diversity of Extrasolar Planets'' (RE 1664/18).
E.R., C.R.-L., M.J.L.-G. and N.M. acknowledge financial support from the Spanish Agencia Estatal de Investigaci\'on through projects AYA2016-79425-C3-3-P, ESP2017-87676-C5-2-R, ESP2017-87143-R and the Centre of Excellence ``Severo Ochoa'' Instituto de Astrof\'isica de Andaluc\'ia (SEV-2017-0709).
J.R.B. and C.A.H. are supported by STFC under grant ST/P000584/1. 
Z.M.B acknowledges CONICYT-FONDECYT/Chile Postdoctorado 3180405. 
E.P. and I.R. acknowledge support from the Spanish Ministry for Science, Innovation and Universities (MCIU) and the Fondo Europeo de Desarrollo Regional (FEDER) through grants ESP2016-80435-C2-1-R and ESP2016-80435-C2-2-R, as well as the support of the Generalitat de Catalunya/CERCA programme. G.A.-E. research is supported by STFC Consolidated Grant ST/P000592/1.

Data were partly obtained with the MONET/South telescope of the MOnitoring NEtwork of Telescopes, funded by the Alfried Krupp von Bohlen und Halbach Foundation, Essen, and operated by the Georg-August-Universit\"at G\"ottingen, the McDonald Observatory of the University of Texas at Austin, and the South African Astronomical Observatory. Data were partly collected with the robotic 40-cm 
telescope ASH2 at the SPACEOBS observatory (San Pedro de Atacama, Chile) operated by the Instituto de Astrof\'\i fica de Andaluc\'\i a (IAA). We acknowledge the effort of various pro-am to obtain photometric monitoring during the Red Dots 2018 campaign through the AAVSO.




\bibliographystyle{mnras}
\bibliography{GJ1061} 

\begin{thebibliography}{}
\makeatletter
\relax
\def\mn@urlcharsother{\let\do\@makeother \do\$\do\&\do\#\do\^\do\_\do\%\do\~}
\def\mn@doi{\begingroup\mn@urlcharsother \@ifnextchar [ {\mn@doi@}
  {\mn@doi@[]}}
\def\mn@doi@[#1]#2{\def\@tempa{#1}\ifx\@tempa\@empty \href
  {http://dx.doi.org/#2} {doi:#2}\else \href {http://dx.doi.org/#2} {#1}\fi
  \endgroup}
\def\mn@eprint#1#2{\mn@eprint@#1:#2::\@nil}
\def\mn@eprint@arXiv#1{\href {http://arxiv.org/abs/#1} {{\tt arXiv:#1}}}
\def\mn@eprint@dblp#1{\href {http://dblp.uni-trier.de/rec/bibtex/#1.xml}
  {dblp:#1}}
\def\mn@eprint@#1:#2:#3:#4\@nil{\def\@tempa {#1}\def\@tempb {#2}\def\@tempc
  {#3}\ifx \@tempc \@empty \let \@tempc \@tempb \let \@tempb \@tempa \fi \ifx
  \@tempb \@empty \def\@tempb {arXiv}\fi \@ifundefined
  {mn@eprint@\@tempb}{\@tempb:\@tempc}{\expandafter \expandafter \csname
  mn@eprint@\@tempb\endcsname \expandafter{\@tempc}}}

\bibitem[\protect\citeauthoryear{{Alibert} \& {Benz}}{{Alibert} \&
  {Benz}}{2017}]{Alibert2017A&A...598L...5A}
{Alibert} Y.,  {Benz} W.,  2017, \mn@doi [\aap] {10.1051/0004-6361/201629671},
  \href {http://adsabs.harvard.edu/abs/2017A%26A...598L...5A} {598, L5}

\bibitem[\protect\citeauthoryear{{Anglada-Escud{\'e}} \&
  {Butler}}{{Anglada-Escud{\'e}} \&
  {Butler}}{2012}]{Anglada2012ApJS..200...15A}
{Anglada-Escud{\'e}} G.,  {Butler} R.~P.,  2012, \mn@doi [\apjs]
  {10.1088/0067-0049/200/2/15}, \href
  {http://adsabs.harvard.edu/abs/2012ApJS..200...15A} {200, 15}

\bibitem[\protect\citeauthoryear{{Anglada-Escud{\'e}}
  et~al.,}{{Anglada-Escud{\'e}} et~al.}{2016}]{Anglada2016Natur.536..437A}
{Anglada-Escud{\'e}} G.,  et~al., 2016, \mn@doi [\nat] {10.1038/nature19106},
  \href {http://adsabs.harvard.edu/abs/2016Natur.536..437A} {536, 437}

\bibitem[\protect\citeauthoryear{{Astudillo-Defru}, {Delfosse}, {Bonfils},
  {Forveille}, {Lovis}  \& {Rameau}}{{Astudillo-Defru}
  et~al.}{2017a}]{Astudillo2017A&A...600A..13A}
{Astudillo-Defru} N.,  {Delfosse} X.,  {Bonfils} X.,  {Forveille} T.,  {Lovis}
  C.,   {Rameau} J.,  2017a, \mn@doi [\aap] {10.1051/0004-6361/201527078},
  \href {http://adsabs.harvard.edu/abs/2017A&A...600A..13A} {600, A13}

\bibitem[\protect\citeauthoryear{{Astudillo-Defru} et~al.,}{{Astudillo-Defru}
  et~al.}{2017b}]{Astudillo2017A&A...602A..88A}
{Astudillo-Defru} N.,  et~al., 2017b, \mn@doi [\aap]
  {10.1051/0004-6361/201630153}, \href
  {http://cdsads.u-strasbg.fr/abs/2017A%26A...602A..88A} {602, A88}

\bibitem[\protect\citeauthoryear{{Astudillo-Defru} et~al.,}{{Astudillo-Defru}
  et~al.}{2017c}]{Astudillo2017A&A...605L..11A}
{Astudillo-Defru} N.,  et~al., 2017c, \mn@doi [\aap]
  {10.1051/0004-6361/201731581}, \href
  {http://cdsads.u-strasbg.fr/abs/2017A%26A...605L..11A} {605, L11}

\bibitem[\protect\citeauthoryear{{Avenhaus}, {Schmid}  \& {Meyer}}{{Avenhaus}
  et~al.}{2012}]{Avenhaus2012A&A...548A.105A}
{Avenhaus} H.,  {Schmid} H.~M.,   {Meyer} M.~R.,  2012, \mn@doi [\aap]
  {10.1051/0004-6361/201219783}, \href
  {http://cdsads.u-strasbg.fr/abs/2012A%26A...548A.105A} {548, A105}

\bibitem[\protect\citeauthoryear{{Baluev}}{{Baluev}}{2009}]{2009MNRAS.393..969B}
{Baluev} R.~V.,  2009, \mn@doi [\mnras] {10.1111/j.1365-2966.2008.14217.x},
  \href {http://adsabs.harvard.edu/abs/2009MNRAS.393..969B} {393, 969}

\bibitem[\protect\citeauthoryear{{Barnes} et~al.,}{{Barnes}
  et~al.}{2014}]{Barnes2014MNRAS.439..3094B}
{Barnes} J.~R.,  et~al., 2014, \mn@doi [\mnras] {10.1093/mnras/stu172}, \href
  {http://adsabs.harvard.edu/abs/2014MNRAS.439.3094B} {439, 3094}

\bibitem[\protect\citeauthoryear{{Barnes}, {Jeffers}, {Haswell}, {Jones},
  {Shulyak}, {Pavlenko}  \& {Jenkins}}{{Barnes}
  et~al.}{2017}]{Barnes2017MNRAS.471..811B}
{Barnes} J.~R.,  {Jeffers} S.~V.,  {Haswell} C.~A.,  {Jones} H.~R.~A.,
  {Shulyak} D.,  {Pavlenko} Y.~V.,   {Jenkins} J.~S.,  2017, \mn@doi [\mnras]
  {10.1093/mnras/stx1482}, \href
  {http://cdsads.u-strasbg.fr/abs/2017MNRAS.471..811B} {471, 811}

\bibitem[\protect\citeauthoryear{{Berta}, {Irwin}, {Charbonneau}, {Burke}  \&
  {Falco}}{{Berta} et~al.}{2012}]{Berta2012AJ....144..145B}
{Berta} Z.~K.,  {Irwin} J.,  {Charbonneau} D.,  {Burke} C.~J.,   {Falco} E.~E.,
   2012, \mn@doi [\aj] {10.1088/0004-6256/144/5/145}, \href
  {http://adsabs.harvard.edu/abs/2012AJ....144..145B} {144, 145}

\bibitem[\protect\citeauthoryear{{Bixel} \& {Apai}}{{Bixel} \&
  {Apai}}{2017}]{bixel2017ApJ...836L..31B}
{Bixel} A.,  {Apai} D.,  2017, \mn@doi [\apjl] {10.3847/2041-8213/aa5f51},
  \href {http://adsabs.harvard.edu/abs/2017ApJ...836L..31B} {836, L31}

\bibitem[\protect\citeauthoryear{{Boro Saikia} et~al.,}{{Boro Saikia}
  et~al.}{2018}]{BoroSakia2018A&A...616A.108B}
{Boro Saikia} S.,  et~al., 2018, \mn@doi [\aap] {10.1051/0004-6361/201629518},
  \href {http://cdsads.u-strasbg.fr/abs/2018A%26A...616A.108B} {616, A108}

\bibitem[\protect\citeauthoryear{{Butler} et~al.,}{{Butler}
  et~al.}{2017}]{2017AJ....153..208B}
{Butler} R.~P.,  et~al., 2017, \mn@doi [\aj] {10.3847/1538-3881/aa66ca}, \href
  {http://adsabs.harvard.edu/abs/2017AJ....153..208B} {153, 208}

\bibitem[\protect\citeauthoryear{{Chambers}}{{Chambers}}{1999}]{Chambers1999MNRAS.304..793C}
{Chambers} J.~E.,  1999, \mn@doi [\mnras] {10.1046/j.1365-8711.1999.02379.x},
  \href {http://adsabs.harvard.edu/abs/1999MNRAS.304..793C} {304, 793}

\bibitem[\protect\citeauthoryear{{Coleman}, {Nelson}, {Paardekooper},
  {Dreizler}, {Giesers}  \& {Anglada-Escud{\'e}}}{{Coleman}
  et~al.}{2017}]{Coleman2017MNRAS.467..996C}
{Coleman} G.~A.~L.,  {Nelson} R.~P.,  {Paardekooper} S.~J.,  {Dreizler} S.,
  {Giesers} B.,   {Anglada-Escud{\'e}} G.,  2017, \mn@doi [\mnras]
  {10.1093/mnras/stx169}, \href
  {http://adsabs.harvard.edu/abs/2017MNRAS.467..996C} {467, 996}

\bibitem[\protect\citeauthoryear{{Coleman}, {Leleu}, {Alibert}  \&
  {Benz}}{{Coleman} et~al.}{2019}]{2019arXiv190804166C}
{Coleman} G. A.~L.,  {Leleu} A.,  {Alibert} Y.,   {Benz} W.,  2019, arXiv
  e-prints, \href {https://ui.adsabs.harvard.edu/abs/2019arXiv190804166C} {p.
  arXiv:1908.04166}

\bibitem[\protect\citeauthoryear{{Damasso} \& {Del Sordo}}{{Damasso} \& {Del
  Sordo}}{2017}]{2017A&A...599A.126D}
{Damasso} M.,  {Del Sordo} F.,  2017, \mn@doi [\aap]
  {10.1051/0004-6361/201630050}, \href
  {https://ui.adsabs.harvard.edu/abs/2017A&A...599A.126D} {599, A126}

\bibitem[\protect\citeauthoryear{Foreman-Mackey}{Foreman-Mackey}{2016}]{corner}
Foreman-Mackey D.,  2016, \mn@doi [The Journal of Open Source Software]
  {10.21105/joss.00024}, 24

\bibitem[\protect\citeauthoryear{{Foreman-Mackey}, {Hogg}, {Lang}  \&
  {Goodman}}{{Foreman-Mackey} et~al.}{2013}]{Foreman-Mackey2013PASP..125..306F}
{Foreman-Mackey} D.,  {Hogg} D.~W.,  {Lang} D.,   {Goodman} J.,  2013, \mn@doi
  [\pasp] {10.1086/670067}, \href
  {http://adsabs.harvard.edu/abs/2013PASP..125..306F} {125, 306}

\bibitem[\protect\citeauthoryear{{Foreman-Mackey}, {Agol}, {Ambikasaran}  \&
  {Angus}}{{Foreman-Mackey} et~al.}{2017}]{2017AJ....154..220F}
{Foreman-Mackey} D.,  {Agol} E.,  {Ambikasaran} S.,   {Angus} R.,  2017,
  \mn@doi [\aj] {10.3847/1538-3881/aa9332}, \href
  {http://adsabs.harvard.edu/abs/2017AJ....154..220F} {154, 220}

\bibitem[\protect\citeauthoryear{{Gaia Collaboration} et~al.,}{{Gaia
  Collaboration} et~al.}{2018}]{GAIA2018A&A...616A...1G}
{Gaia Collaboration} et~al., 2018, \mn@doi [\aap]
  {10.1051/0004-6361/201833051}, \href
  {http://adsabs.harvard.edu/abs/2018A%26A...616A...1G} {616, A1}

\bibitem[\protect\citeauthoryear{{Gaidos} et~al.,}{{Gaidos}
  et~al.}{2014}]{Gaidos2014MNRAS.443.2561G}
{Gaidos} E.,  et~al., 2014, \mn@doi [\mnras] {10.1093/mnras/stu1313}, \href
  {http://cdsads.u-strasbg.fr/abs/2014MNRAS.443.2561G} {443, 2561}

\bibitem[\protect\citeauthoryear{{Gillon} et~al.,}{{Gillon}
  et~al.}{2017}]{Gillon2017Natur.542..456G}
{Gillon} M.,  et~al., 2017, \mn@doi [\nat] {10.1038/nature21360}, \href
  {http://adsabs.harvard.edu/abs/2017Natur.542..456G} {542, 456}

\bibitem[\protect\citeauthoryear{{Grimm} et~al.,}{{Grimm}
  et~al.}{2018}]{Grimm2018A&A...613A..68G}
{Grimm} S.~L.,  et~al., 2018, \mn@doi [\aap] {10.1051/0004-6361/201732233},
  \href {http://adsabs.harvard.edu/abs/2018A%26A...613A..68G} {613, A68}

\bibitem[\protect\citeauthoryear{{Hambsch}}{{Hambsch}}{2012}]{2012SASS...31...75H}
{Hambsch} F.-J.,  2012, Society for Astronomical Sciences Annual Symposium,
  \href {https://ui.adsabs.harvard.edu/abs/2012SASS...31...75H} {31, 75}

\bibitem[\protect\citeauthoryear{{Henry}, {Walkowicz}, {Barto}  \&
  {Golimowski}}{{Henry} et~al.}{2002}]{Henry2002AJ....123.2002H}
{Henry} T.~J.,  {Walkowicz} L.~M.,  {Barto} T.~C.,   {Golimowski} D.~A.,  2002,
  \mn@doi [\aj] {10.1086/339315}, \href
  {http://adsabs.harvard.edu/abs/2002AJ....123.2002H} {123, 2002}

\bibitem[\protect\citeauthoryear{{Jeffers} et~al.,}{{Jeffers}
  et~al.}{2018a}]{2018A&A...614A..76J}
{Jeffers} S.~V.,  et~al., 2018a, \mn@doi [\aap] {10.1051/0004-6361/201629599},
  \href {https://ui.adsabs.harvard.edu/abs/2018A&A...614A..76J} {614, A76}

\bibitem[\protect\citeauthoryear{{Jeffers} et~al.,}{{Jeffers}
  et~al.}{2018b}]{Jeffers2018A&A...614A..76J}
{Jeffers} S.~V.,  et~al., 2018b, \mn@doi [\aap] {10.1051/0004-6361/201629599},
  \href {http://adsabs.harvard.edu/abs/2018A%26A...614A..76J} {614, A76}

\bibitem[\protect\citeauthoryear{{Kochanek} et~al.,}{{Kochanek}
  et~al.}{2017}]{Kochanek2017PASP..129j4502K}
{Kochanek} C.~S.,  et~al., 2017, \mn@doi [\pasp] {10.1088/1538-3873/aa80d9},
  \href {http://adsabs.harvard.edu/abs/2017PASP..129j4502K} {129, 104502}

\bibitem[\protect\citeauthoryear{{Kopparapu}, {Ramirez}, {SchottelKotte},
  {Kasting}, {Domagal-Goldman}  \& {Eymet}}{{Kopparapu}
  et~al.}{2014}]{Kopparapu2014ApJ...787L..29K}
{Kopparapu} R.~K.,  {Ramirez} R.~M.,  {SchottelKotte} J.,  {Kasting} J.~F.,
  {Domagal-Goldman} S.,   {Eymet} V.,  2014, \mn@doi [\apjl]
  {10.1088/2041-8205/787/2/L29}, \href
  {http://adsabs.harvard.edu/abs/2014ApJ...787L..29K} {787, L29}

\bibitem[\protect\citeauthoryear{{Lo Curto} et~al.,}{{Lo Curto}
  et~al.}{2015}]{LoCurto2015Msngr.162....9L}
{Lo Curto} G.,  et~al., 2015, The Messenger, \href
  {http://cdsads.u-strasbg.fr/abs/2015Msngr.162....9L} {162, 9}

\bibitem[\protect\citeauthoryear{{Luque} et~al.,}{{Luque}
  et~al.}{2019}]{2019arXiv190412818L}
{Luque} R.,  et~al., 2019, arXiv e-prints, \href
  {https://ui.adsabs.harvard.edu/abs/2019arXiv190412818L} {p. arXiv:1904.12818}

\bibitem[\protect\citeauthoryear{{Mayor} et~al.,}{{Mayor}
  et~al.}{2003}]{Mayor2003Msngr.114...20M}
{Mayor} M.,  et~al., 2003, The Messenger, \href
  {http://cdsads.u-strasbg.fr/abs/2003Msngr.114...20M} {114, 20}

\bibitem[\protect\citeauthoryear{{Meschiari}, {Wolf}, {Rivera}, {Laughlin},
  {Vogt}  \& {Butler}}{{Meschiari} et~al.}{2009}]{Meschiari2009PASP..121.1016M}
{Meschiari} S.,  {Wolf} A.~S.,  {Rivera} E.,  {Laughlin} G.,  {Vogt} S.,
  {Butler} P.,  2009, \mn@doi [\pasp] {10.1086/605730}, \href
  {http://adsabs.harvard.edu/abs/2009PASP..121.1016M} {121, 1016}

\bibitem[\protect\citeauthoryear{{Mortier} \& {Collier Cameron}}{{Mortier} \&
  {Collier Cameron}}{2017}]{2017A&A...601A.110M}
{Mortier} A.,  {Collier Cameron} A.,  2017, \mn@doi [\aap]
  {10.1051/0004-6361/201630201}, \href
  {https://ui.adsabs.harvard.edu/abs/2017A&A...601A.110M} {601, A110}

\bibitem[\protect\citeauthoryear{{Mortier}, {Faria}, {Correia}, {Santerne}  \&
  {Santos}}{{Mortier} et~al.}{2015}]{2015A&A...573A.101M}
{Mortier} A.,  {Faria} J.~P.,  {Correia} C.~M.,  {Santerne} A.,   {Santos}
  N.~C.,  2015, \mn@doi [\aap] {10.1051/0004-6361/201424908}, \href
  {https://ui.adsabs.harvard.edu/abs/2015A&A...573A.101M} {573, A101}

\bibitem[\protect\citeauthoryear{{Neves}, {Bonfils}, {Santos}, {Delfosse},
  {Forveille}, {Allard}  \& {Udry}}{{Neves}
  et~al.}{2014}]{Neves2014A&A...568A.121N}
{Neves} V.,  {Bonfils} X.,  {Santos} N.~C.,  {Delfosse} X.,  {Forveille} T.,
  {Allard} F.,   {Udry} S.,  2014, \mn@doi [\aap]
  {10.1051/0004-6361/201424139}, \href
  {http://adsabs.harvard.edu/abs/2014A%26A...568A.121N} {568, A121}

\bibitem[\protect\citeauthoryear{{Newton}, {Irwin}, {Charbonneau}, {Berlind},
  {Calkins}  \& {Mink}}{{Newton} et~al.}{2017}]{Newton2017ApJ...834...85N}
{Newton} E.~R.,  {Irwin} J.,  {Charbonneau} D.,  {Berlind} P.,  {Calkins}
  M.~L.,   {Mink} J.,  2017, \mn@doi [\apj] {10.3847/1538-4357/834/1/85}, \href
  {http://adsabs.harvard.edu/abs/2017ApJ...834...85N} {834, 85}

\bibitem[\protect\citeauthoryear{{Pizzolato}, {Maggio}, {Micela}, {Sciortino}
  \& {Ventura}}{{Pizzolato} et~al.}{2003}]{2003A&A...397..147P}
{Pizzolato} N.,  {Maggio} A.,  {Micela} G.,  {Sciortino} S.,   {Ventura} P.,
  2003, \mn@doi [\aap] {10.1051/0004-6361:20021560}, \href
  {http://adsabs.harvard.edu/abs/2003A%26A...397..147P} {397, 147}

\bibitem[\protect\citeauthoryear{{Reiners}, {Joshi}  \& {Goldman}}{{Reiners}
  et~al.}{2012}]{Reiners2012AJ....143...93R}
{Reiners} A.,  {Joshi} N.,   {Goldman} B.,  2012, \mn@doi [\aj]
  {10.1088/0004-6256/143/4/93}, \href
  {https://ui.adsabs.harvard.edu/\#abs/2012AJ....143...93R} {143, 93}

\bibitem[\protect\citeauthoryear{{Reiners}, {Sch{\"u}ssler}  \&
  {Passegger}}{{Reiners} et~al.}{2014}]{Reiners2014ApJ...794..144R}
{Reiners} A.,  {Sch{\"u}ssler} M.,   {Passegger} V.~M.,  2014, \mn@doi [\apj]
  {10.1088/0004-637X/794/2/144}, \href
  {http://adsabs.harvard.edu/abs/2014ApJ...794..144R} {794, 144}

\bibitem[\protect\citeauthoryear{{Ribas} et~al.,}{{Ribas}
  et~al.}{2018}]{Ribas2018Natur.563..365R}
{Ribas} I.,  et~al., 2018, \mn@doi [\nat] {10.1038/s41586-018-0677-y}, \href
  {http://adsabs.harvard.edu/abs/2018Natur.563..365R} {563, 365}

\bibitem[\protect\citeauthoryear{{Ricker} et~al.,}{{Ricker}
  et~al.}{2015a}]{2015JATIS...1a4003R}
{Ricker} G.~R.,  et~al., 2015a, \mn@doi [Journal of Astronomical Telescopes,
  Instruments, and Systems] {10.1117/1.JATIS.1.1.014003}, \href
  {https://ui.adsabs.harvard.edu/abs/2015JATIS...1a4003R} {1, 014003}

\bibitem[\protect\citeauthoryear{{Ricker} et~al.,}{{Ricker}
  et~al.}{2015b}]{Ricker2015JATIS...1a4003R}
{Ricker} G.~R.,  et~al., 2015b, \mn@doi [Journal of Astronomical Telescopes,
  Instruments, and Systems] {10.1117/1.JATIS.1.1.014003}, \href
  {http://adsabs.harvard.edu/abs/2015JATIS...1a4003R} {1, 014003}

\bibitem[\protect\citeauthoryear{{Rivera}, {Laughlin}, {Butler}, {Vogt},
  {Haghighipour}  \& {Meschiari}}{{Rivera}
  et~al.}{2010}]{Rivera2010ApJ...719..890R}
{Rivera} E.~J.,  {Laughlin} G.,  {Butler} R.~P.,  {Vogt} S.~S.,  {Haghighipour}
  N.,   {Meschiari} S.,  2010, \mn@doi [\apj] {10.1088/0004-637X/719/1/890},
  \href {http://cdsads.u-strasbg.fr/abs/2010ApJ...719..890R} {719, 890}

\bibitem[\protect\citeauthoryear{{Rodgers} \& {Eggen}}{{Rodgers} \&
  {Eggen}}{1974}]{Rodgers1974PASP...86..742R}
{Rodgers} A.~W.,  {Eggen} O.~J.,  1974, \mn@doi [\pasp] {10.1086/129670}, \href
  {http://adsabs.harvard.edu/abs/1974PASP...86..742R} {86, 742}

\bibitem[\protect\citeauthoryear{{Rodriguez}, {Duch{\^e}ne}, {Tom}, {Kennedy},
  {Matthews}, {Greaves}  \& {Butner}}{{Rodriguez}
  et~al.}{2015}]{Rodriguez2015MNRAS.449.3160R}
{Rodriguez} D.~R.,  {Duch{\^e}ne} G.,  {Tom} H.,  {Kennedy} G.~M.,  {Matthews}
  B.,  {Greaves} J.,   {Butner} H.,  2015, \mn@doi [\mnras]
  {10.1093/mnras/stv483}, \href
  {http://cdsads.u-strasbg.fr/abs/2015MNRAS.449.3160R} {449, 3160}

\bibitem[\protect\citeauthoryear{{Schmitt} \& {Liefke}}{{Schmitt} \&
  {Liefke}}{2004}]{Schmitt2004A&A...417..651S}
{Schmitt} J.~H.~M.~M.,  {Liefke} C.,  2004, \mn@doi [\aap]
  {10.1051/0004-6361:20030495}, \href
  {http://cdsads.u-strasbg.fr/abs/2004A%26A...417..651S} {417, 651}

\bibitem[\protect\citeauthoryear{{Skrutskie} et~al.,}{{Skrutskie}
  et~al.}{2006}]{Skrutskie2006AJ....131.1163S}
{Skrutskie} M.~F.,  et~al., 2006, \mn@doi [\aj] {10.1086/498708}, \href
  {http://adsabs.harvard.edu/abs/2006AJ....131.1163S} {131, 1163}

\bibitem[\protect\citeauthoryear{{Stassun} et~al.,}{{Stassun}
  et~al.}{2018}]{2018AJ....156..102S}
{Stassun} K.~G.,  et~al., 2018, \mn@doi [\aj] {10.3847/1538-3881/aad050}, \href
  {http://adsabs.harvard.edu/abs/2018AJ....156..102S} {156, 102}

\bibitem[\protect\citeauthoryear{{Su{\'a}rez Mascare{\~n}o}, {Rebolo},
  {Gonz{\'a}lez Hern{\'a}ndez}  \& {Esposito}}{{Su{\'a}rez Mascare{\~n}o}
  et~al.}{2015}]{2015MNRAS.452.2745S}
{Su{\'a}rez Mascare{\~n}o} A.,  {Rebolo} R.,  {Gonz{\'a}lez Hern{\'a}ndez}
  J.~I.,   {Esposito} M.,  2015, \mn@doi [\mnras] {10.1093/mnras/stv1441},
  \href {https://ui.adsabs.harvard.edu/abs/2015MNRAS.452.2745S} {452, 2745}

\bibitem[\protect\citeauthoryear{{West}, {Hawley}, {Bochanski}, {Covey},
  {Reid}, {Dhital}, {Hilton}  \& {Masuda}}{{West}
  et~al.}{2008}]{West2008AJ....135..785W}
{West} A.~A.,  {Hawley} S.~L.,  {Bochanski} J.~J.,  {Covey} K.~R.,  {Reid}
  I.~N.,  {Dhital} S.,  {Hilton} E.~J.,   {Masuda} M.,  2008, \mn@doi [\aj]
  {10.1088/0004-6256/135/3/785}, \href
  {http://adsabs.harvard.edu/abs/2008AJ....135..785W} {135, 785}

\bibitem[\protect\citeauthoryear{{Wright}, {Drake}, {Mamajek}  \&
  {Henry}}{{Wright} et~al.}{2011}]{Wright2011ApJ...743...48W}
{Wright} N.~J.,  {Drake} J.~J.,  {Mamajek} E.~E.,   {Henry} G.~W.,  2011,
  \mn@doi [\apj] {10.1088/0004-637X/743/1/48}, \href
  {http://adsabs.harvard.edu/abs/2011ApJ...743...48W} {743, 48}

\bibitem[\protect\citeauthoryear{{Wright}, {Wittenmyer}, {Tinney}, {Bentley}
  \& {Zhao}}{{Wright} et~al.}{2016}]{Wright2016ApJ...817L..20W}
{Wright} D.~J.,  {Wittenmyer} R.~A.,  {Tinney} C.~G.,  {Bentley} J.~S.,
  {Zhao} J.,  2016, \mn@doi [\apjl] {10.3847/2041-8205/817/2/L20}, \href
  {http://cdsads.u-strasbg.fr/abs/2016ApJ...817L..20W} {817, L20}

\bibitem[\protect\citeauthoryear{{Zechmeister} \& {K{\"u}rster}}{{Zechmeister}
  \& {K{\"u}rster}}{2009}]{Zechmeister2009A&A...496..577Z}
{Zechmeister} M.,  {K{\"u}rster} M.,  2009, \mn@doi [\aap]
  {10.1051/0004-6361:200811296}, \href
  {http://adsabs.harvard.edu/abs/2009A%26A...496..577Z} {496, 577}

\bibitem[\protect\citeauthoryear{{Zechmeister} et~al.,}{{Zechmeister}
  et~al.}{2018}]{Zechmeister2018A&A...609A..12Z}
{Zechmeister} M.,  et~al., 2018, \mn@doi [\aap] {10.1051/0004-6361/201731483},
  \href {http://adsabs.harvard.edu/abs/2018A%26A...609A..12Z} {609, A12}

\bibitem[\protect\citeauthoryear{{Zechmeister} et~al.,}{{Zechmeister}
  et~al.}{2019}]{2019A&A...627A..49Z}
{Zechmeister} M.,  et~al., 2019, \mn@doi [\aap] {10.1051/0004-6361/201935460},
  \href {https://ui.adsabs.harvard.edu/abs/2019A&A...627A..49Z} {627, A49}

\bibitem[\protect\citeauthoryear{{Zeng}, {Sasselov}  \& {Jacobsen}}{{Zeng}
  et~al.}{2016}]{Zeng2016ApJ...819..127Z}
{Zeng} L.,  {Sasselov} D.~D.,   {Jacobsen} S.~B.,  2016, \mn@doi [\apj]
  {10.3847/0004-637X/819/2/127}, \href
  {http://adsabs.harvard.edu/abs/2016ApJ...819..127Z} {819, 127}

\makeatother
\end{thebibliography}




\appendix
\section{Radial Velocity data}
In  Table\,\ref{tab:RV Hpre} and \ref{tab:RV Hpost} we list the all radial velocity data for GJ\,1061. Radial velocities are obtained using {\em  SERVAL} and {\em  TERRA}. Note that the full Table\,\ref{tab:RV Hpost} is available online.

\begin{table}
   \centering
   \caption{\label{tab:RV Hpre} HARPS pre-upgrade, pre-Red Dots data obtained by {\em SERVAL}.}
   \begin{tabular}{crr}
     \hline 
     \hline
      BJD & RV & $\sigma_{\rm RV}$ \\
          & [m/s] & [m/s] \\
\hline
2452985.714000	&	-35.96	&	1.82	\\
2452996.738000	&	-25.68	&	2.64	\\
2453337.750000	&	-42.07	&	1.64	\\
2454341.869000	&	-15.33	&	2.04	\\
2455545.668000	&	-33.95	&	1.63	\\
2455612.524000	&	-13.61	&	1.83	\\
2455998.513000	&	-9.65	&	1.72	\\

      \hline
    \end{tabular}
\end{table}

\begin{table}
   \centering
   \caption{\label{tab:RV Hpost}\label{tab:RV PRD} HARPS post-upgrade data obtained by {\em TERRA} and {\em SERVAL} (dLW). Outliers are flagged (RV or dLW) in the last column. Data are given in 1-day averages and 1\,m/s has been quadratically added to the RV uncertainties. Full table in online data.} 
   \begin{tabular}{crrrrr}
     \hline 
     \hline
      BJD & RV & $\sigma_{\rm RV}$ & dLW &  $\sigma_{\rm dLW}$ & flag\\
          & [m/s] & [m/s]  & [${\rm m^2/s^2}$] & [${\rm m^2/s^2}$] & \\
\hline
2458020.814834 & 1.48 & 1.69 & -12.1 & 3.6 \\
2458040.854023 & -6.77 & 1.06 & 2.9 & 1.4 \\
2458043.783336 & 2.56 & 1.66 & 0.9 & 1.4 \\
2458052.765154 & -1.54 & 1.59 & 2.6 & 1.4 \\
2458053.866459 & -6.16 & 1.05 & 7.7 & 1.4 \\
2458054.699823 & -7.11 & 1.26 & -0.3 & 1.2 \\
       \hline
    \end{tabular}
\end{table}

\section{Modeling details}
\label{App:models}
 We start our {\em python}-based detailed analysis using {\em scipy.optimize.minimize}. For each Keplerian signal we use the parameters provided from the GLS analysis as starting values (amplitude, period, periastron time). The best fit and its uncertainties are used to then initialize the MCMC run with 500 walkers and 3000 steps. As priors for all parameters, we use a normal distribution with the best fit as mean and 100 times the uncertainty as the standard deviation. After the run we check that the parameter distribution was not affected by the initialisation. Boundaries are set only for numerical reasons, e.g. to avoid numerical problems with too large or too small values for the kernel parameters. To avoid non-physical parameter combinations, the absolute value of the eccentricity is used when calculating the Keplerian models. The absolute value of the velocity amplitude is used in order to avoid degenerate solutions. In case the final best solution from the MCMC chains is significantly better than the starting parameters, this procedure is repeated with the new best parameter set.

The correlation of the data is modeled using the Gaussian Process Regression framework. We use the {\em celerite} package \citep{2017AJ....154..220F} since it only scales linearly with the number of data points and not cubical.  
The {\em REAL} kernel of {\em celerite} represents an exponential decaying coherence of the data and has two parameters, the amplitude $a$ and the inverse damping time scale $c$ (to avoid negative values for $a$ and $c$, the logarithm of the values are used):
\begin{displaymath}
k(\tau) = a\,e^{-c\,\tau}
\end{displaymath}

The {\em SHO} kernel represents a stochastically driven, damped harmonic oscillator. It is defined through its power spectral density (PSD) using three parameters, an amplitude $S_0$, the quality factor $Q$ describing the damping and the oscillator frequency $\omega_0$. Like in the {\em REAL} kernel, the logarithms of the parameters are used. The quality factor is related to the inverse of the damping time. The PSD of this term is:
\begin{displaymath}
S(\omega) = \sqrt{\frac{2}{\pi}} \frac{S_0\,\omega_0^4}
{(\omega^2-{\omega_0}^2)^2 + {\omega_0}^2\,\omega^2/Q^2}
\end{displaymath}

\section{\label{sec:Corner}MCMC corner plots}
\begin{figure*}
    \centering
    \includegraphics[width=\textwidth]{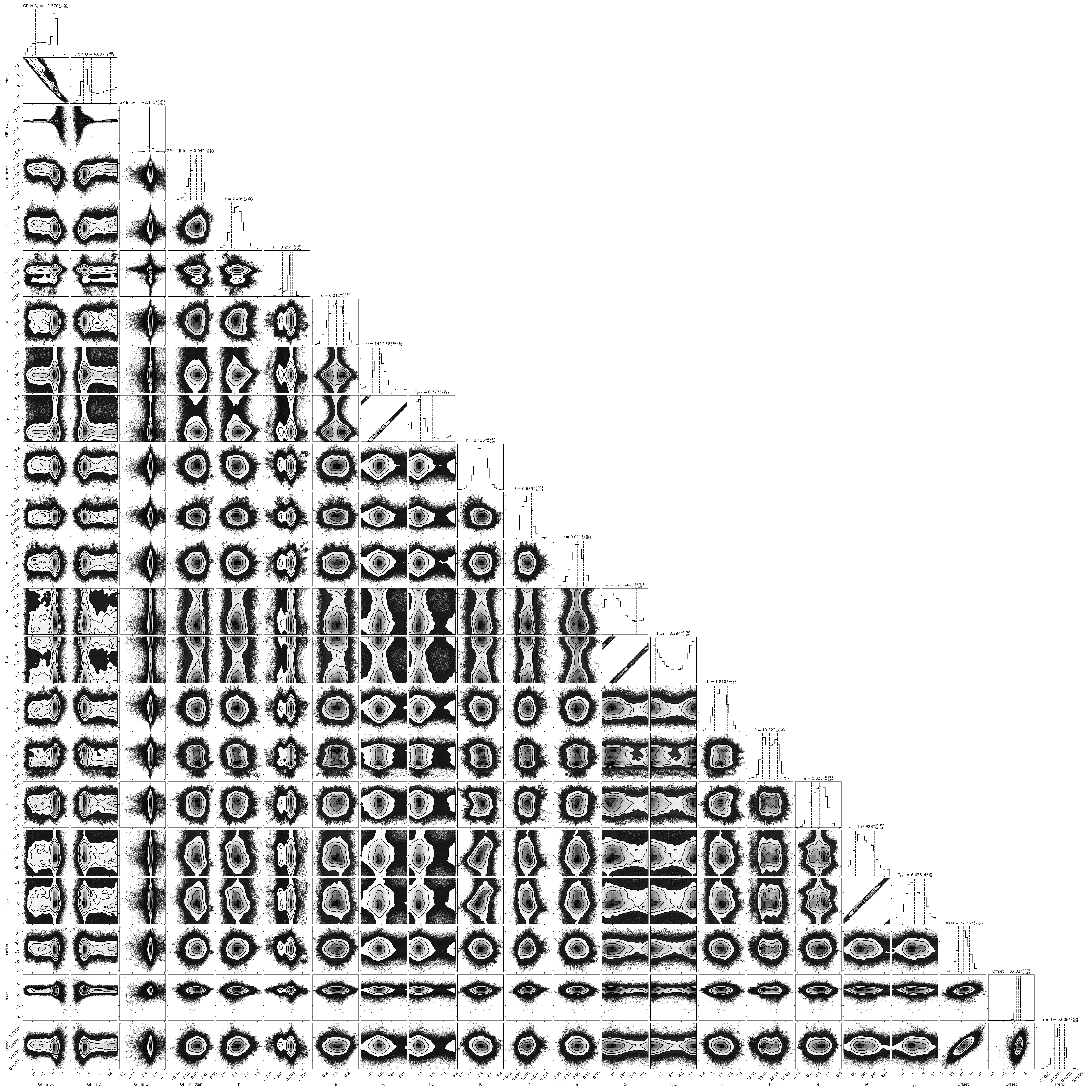}
    \caption{MCMC posterior distribution of full parameter set for the 3-Keplerian signals plus {\em SHO} model.}
    \label{fig:MCMC_full}
\end{figure*}
\begin{figure*}
   \centering
   \includegraphics[width=0.49\linewidth]{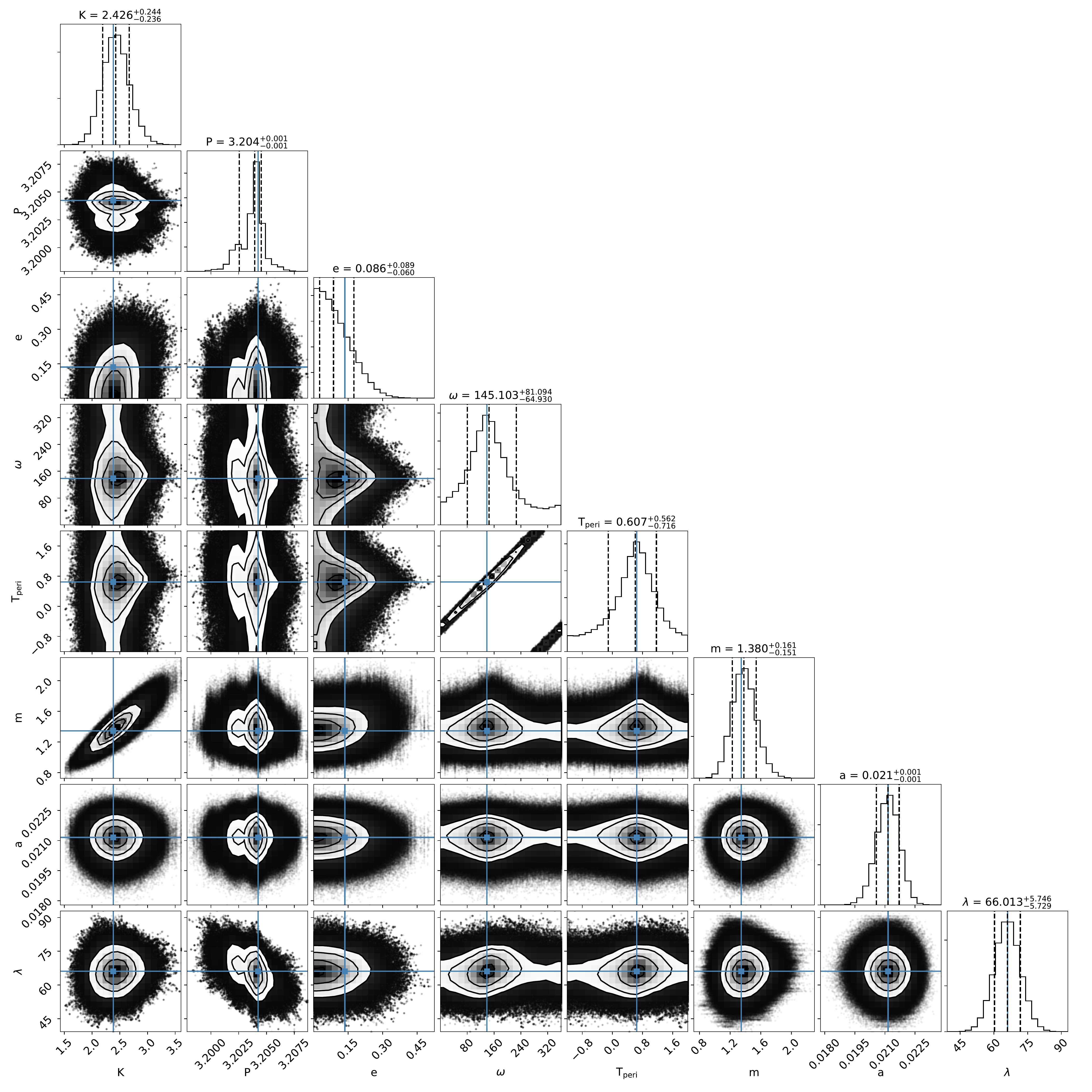}
   \includegraphics[width=0.49\linewidth]{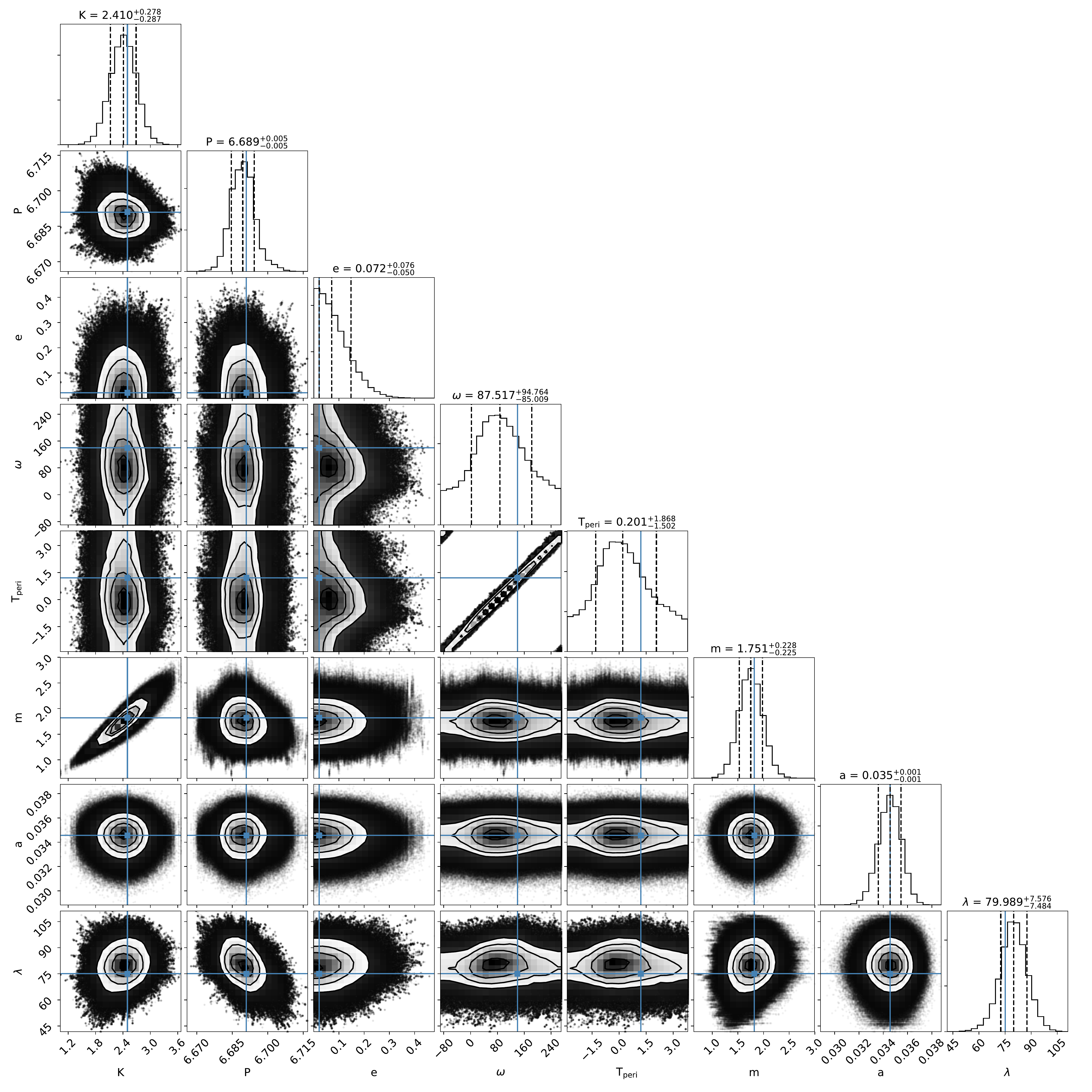}
   \caption{\label{fig:MCMC_b}MCMC posterior distribution for Keplerian parameters of planet b and c.}
\end{figure*}
\begin{figure*}
   \centering
   \includegraphics[width=0.49\linewidth]{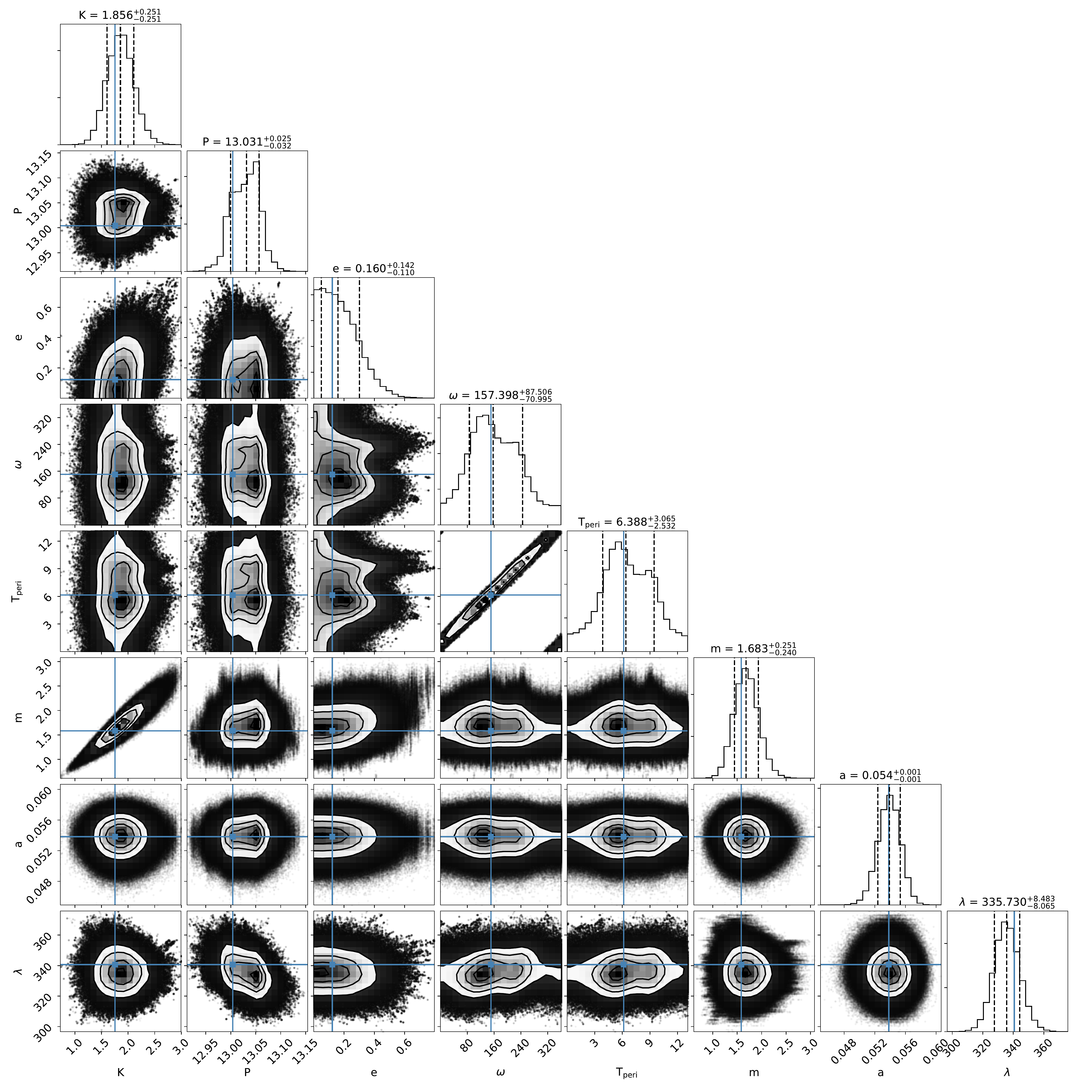}
   \includegraphics[width=0.49\linewidth]{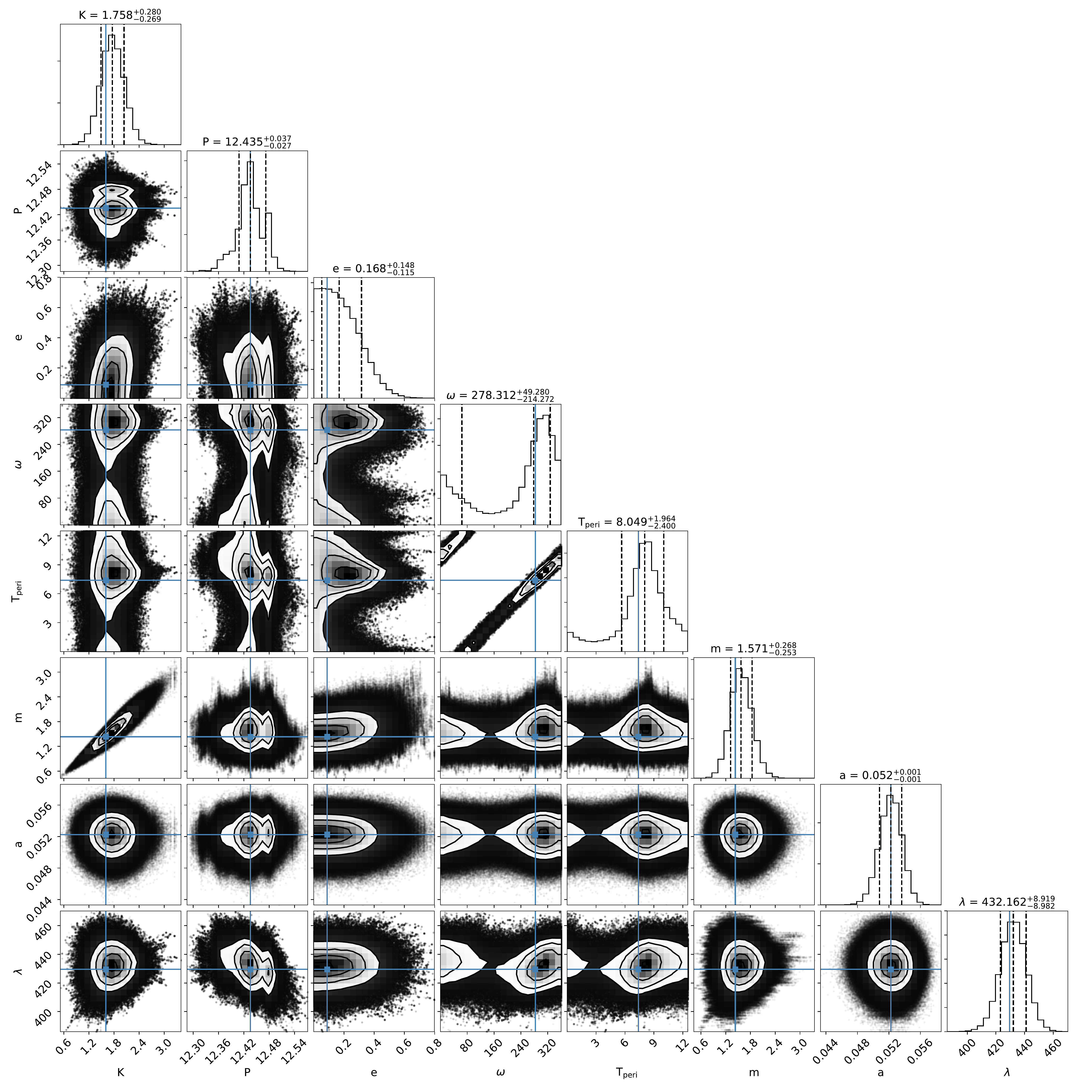}
   \caption{\label{fig:MCMC_d}Same as Fig.~\ref{fig:MCMC_b} but for planet d choosing 13.0 and 12.4\,d period.}
\end{figure*}
\begin{figure}
   \centering
   \includegraphics[width=0.99\linewidth]{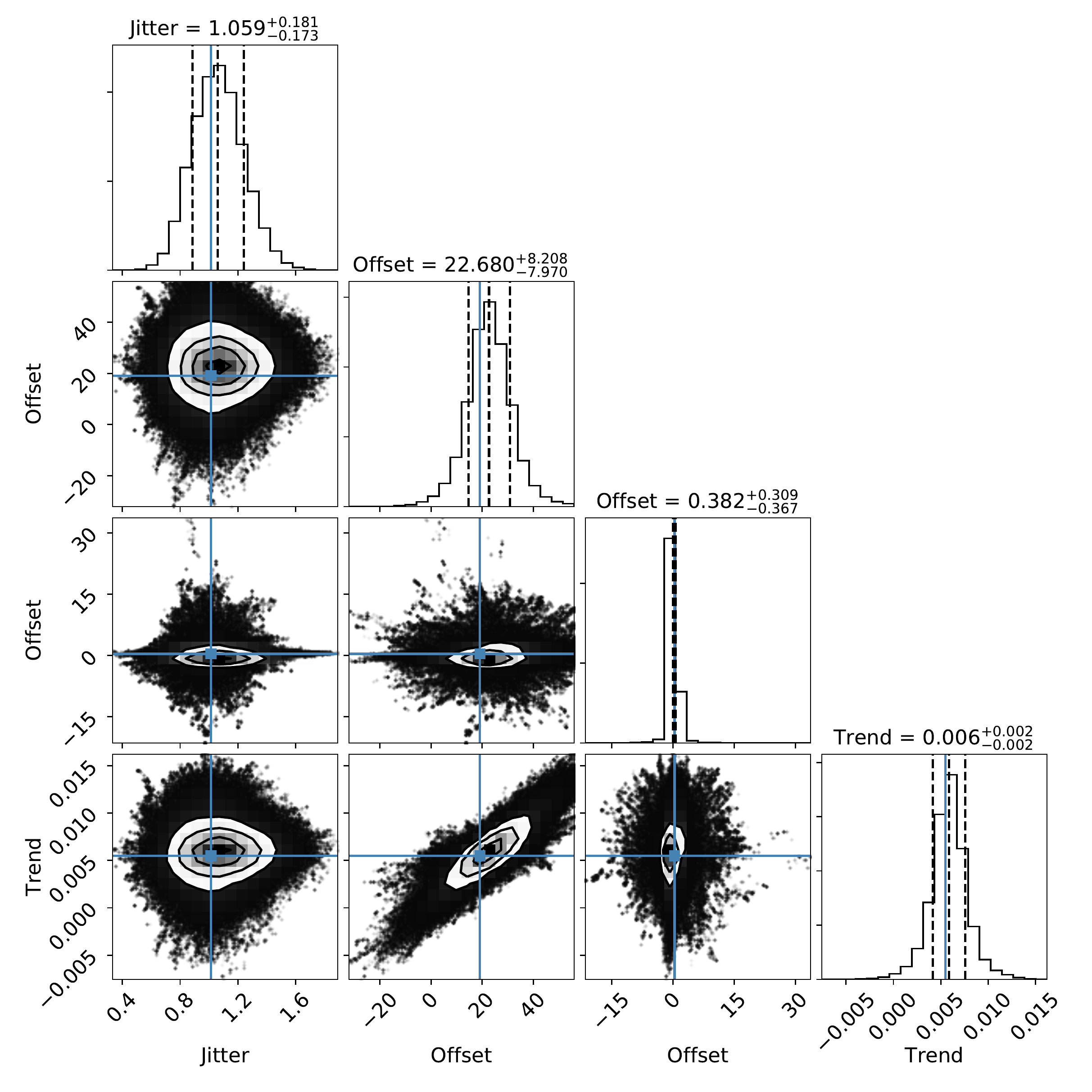}
   \caption{\label{fig:MCMC_jitter}{\em Left}: MCMC posterior distribution for $\ln (\mathrm{jitter})$, offsets, and trend parameters. {\em Right}: Parameter for the {\em SHO} kernel.}
\end{figure}

 We show the posterior distribution from 500 walkers and 30\,000 steps (the first half rejected as burn-in) are as corner plots using {\em corner} \citep{corner} in  Figs.\,\ref{fig:MCMC_full} to \ref{fig:MCMC_jitter}, first for the full fit parameter set and then parameter subsets for the individual planets (Figs.\,\ref{fig:MCMC_b} and \ref{fig:MCMC_d}) as well as for instrumental parameters (Fig.\,\,\ref{fig:MCMC_jitter}). For the parameter subsets we show absolute values for the eccentricities, the longitude of periastron and the periastron passage time are converted modulo $2\pi$ and orbital period, respectively. The subset corner plots also contain derived parameters, i.e. the semi-major axis, the planetary minimum mass as well as the longitude of periastron. For the former two we propagate the uncertainties in the stellar mass. The 4$^{\mathrm{th}}$ signal is modeled using the {\em SHO}-kernel.



\section{Periodograms}
\label{app:periodogram}

\begin{figure}
   \centering
   \includegraphics[width=0.52\textwidth]{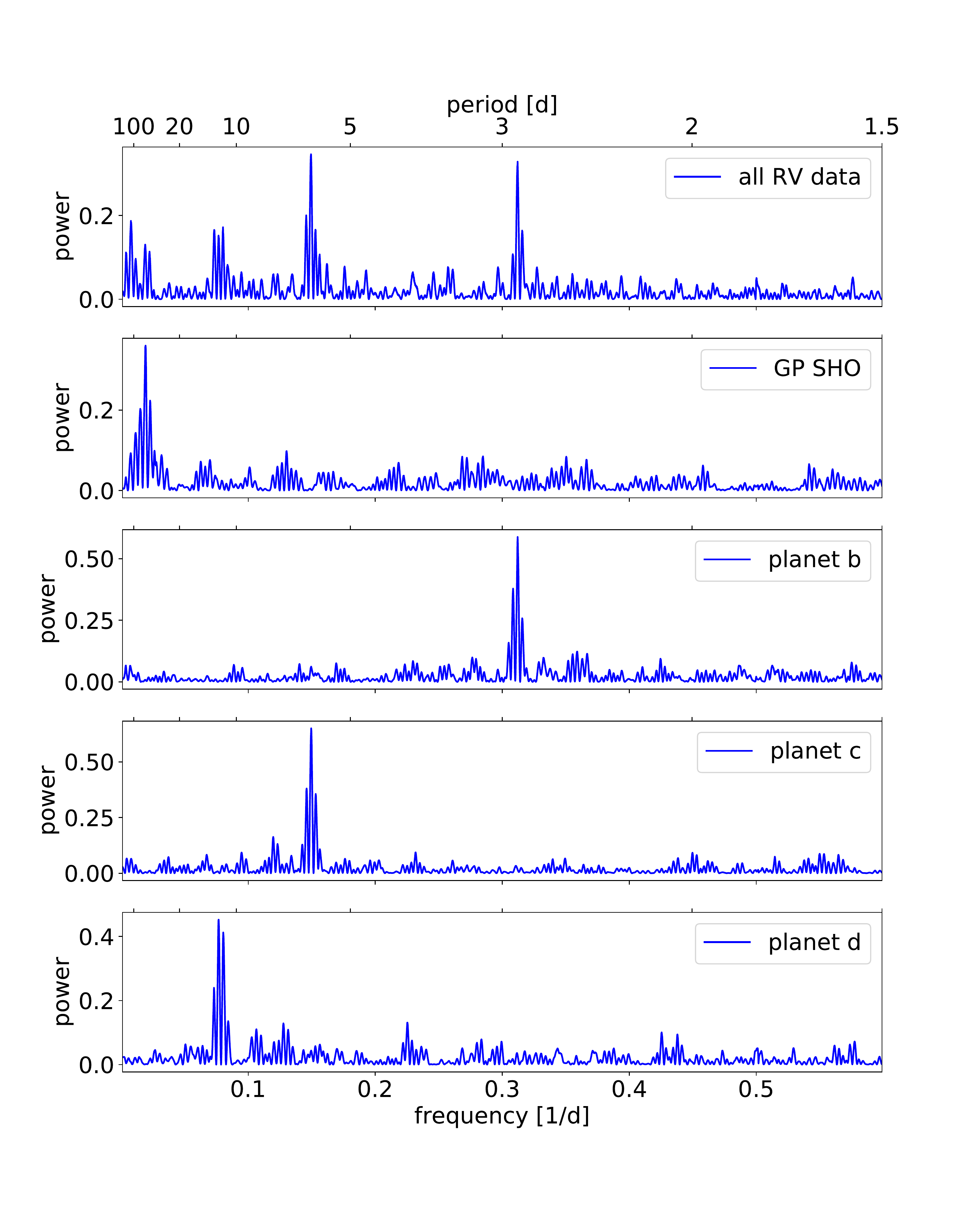}
   \caption{\label{fig:AllRV_signalsapp} Periodograms of all observation (top) and models.}
\end{figure}

 The subsequent subtraction of Keplerian signals from the radial velocity data is shown in Fig.\ref{fig:AllRV_prewhite} as GLS-periodograms \citep{Zechmeister2009A&A...496..577Z}. In addition, the periodograms of the signals are shown in Fig.\ref{fig:AllRV_signalsapp}. 

\begin{figure}
    \centering
    \includegraphics[width=1\linewidth]{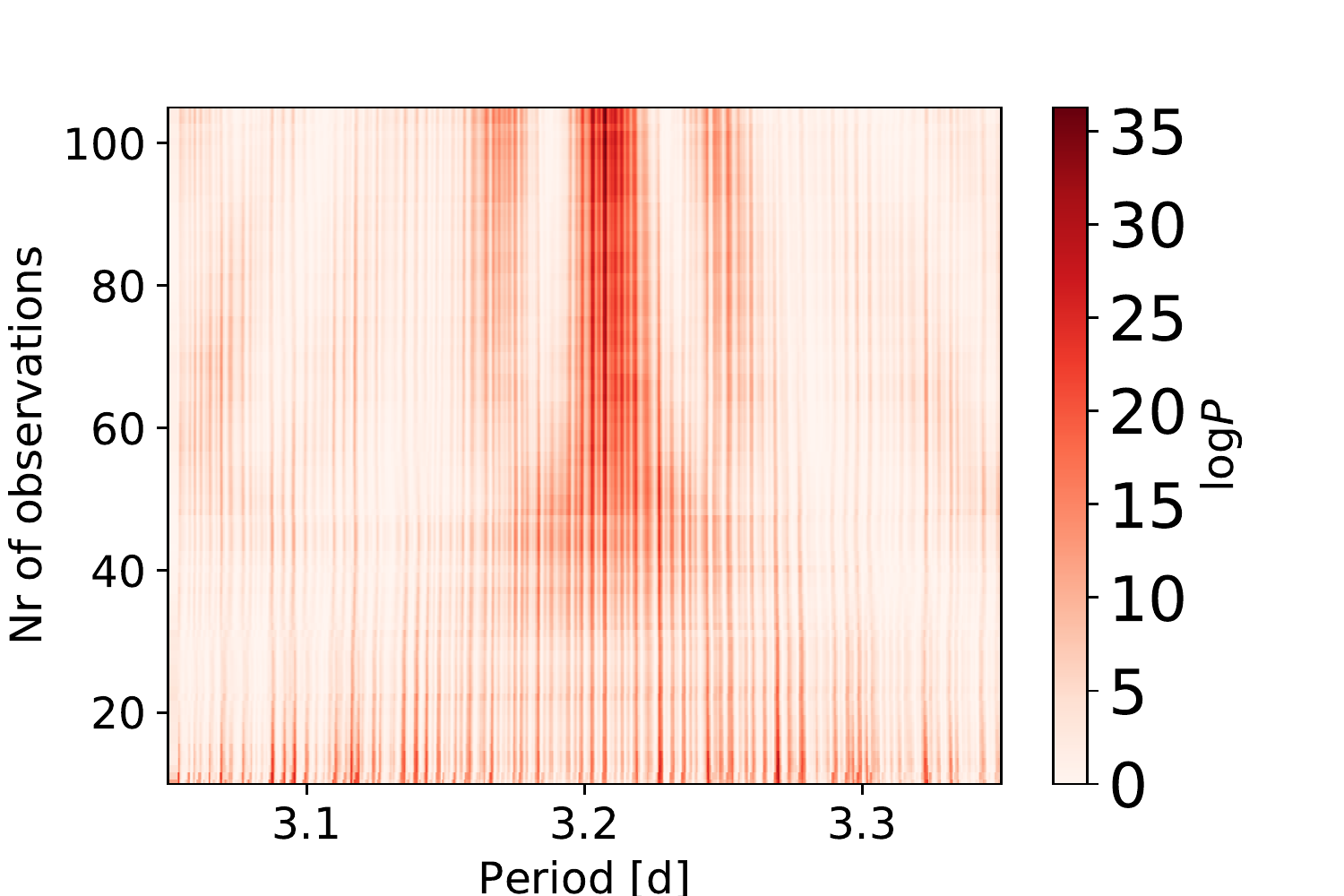}
    \includegraphics[width=1\linewidth]{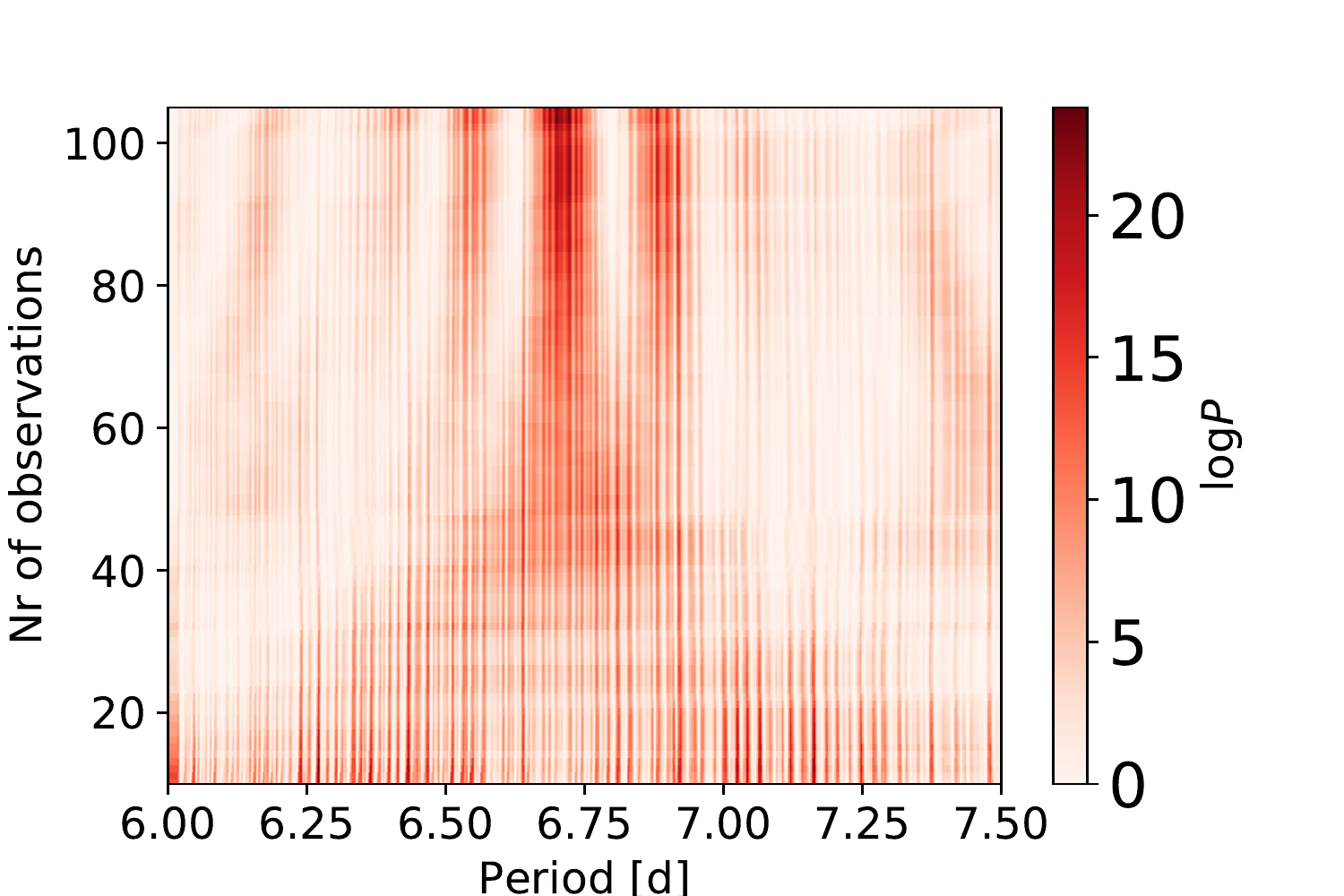}
    \includegraphics[width=1\linewidth]{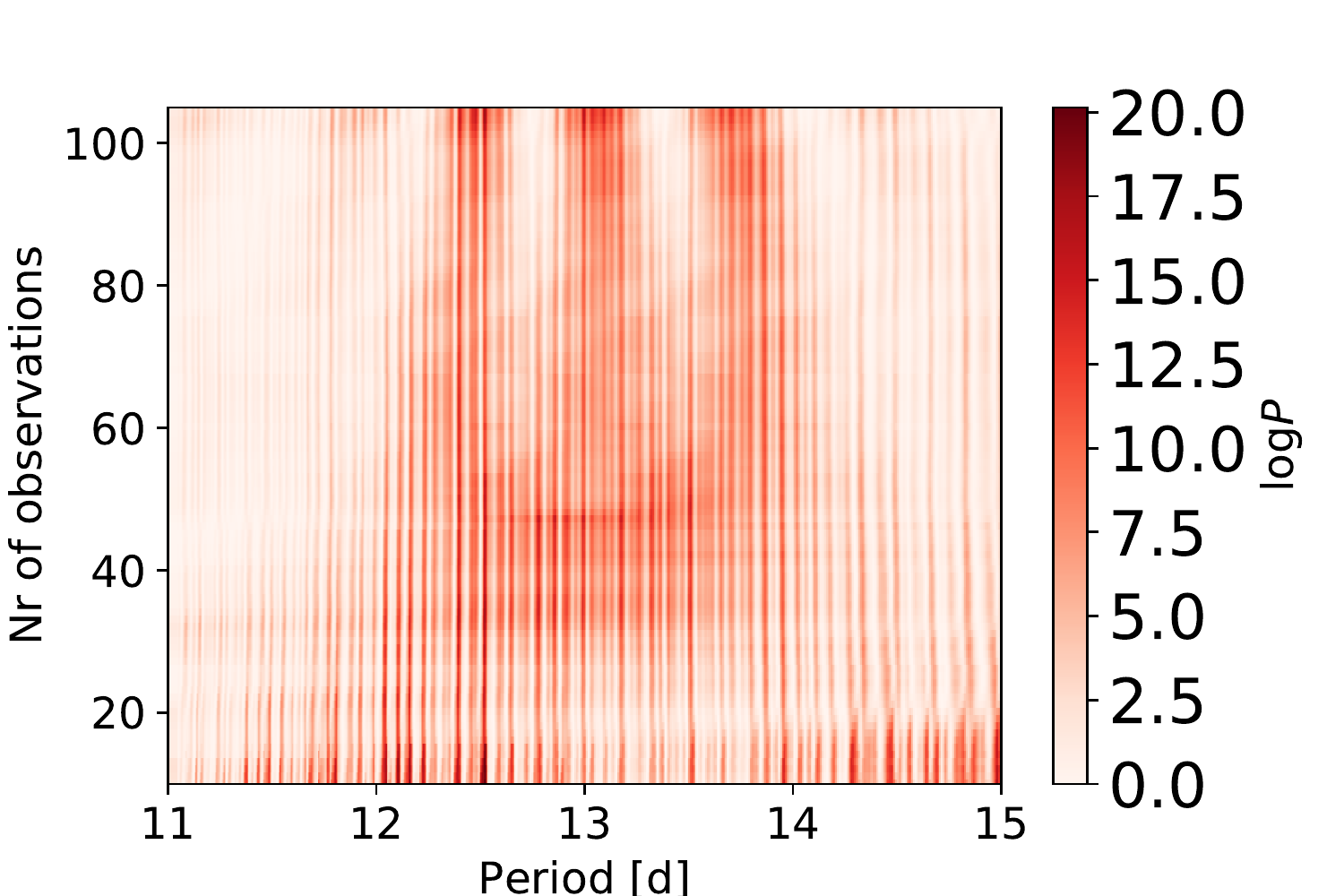}
    \caption{SBGLS periodograms zoomed around the 3.204\,d period (top) and subsequent removal of the 3.204\,d (middle) as well as the 6.689\,d signal (bottom). The number of observations is plotted against period, with the colour scale indicating the logarithm of the probability, where darker is more likely.}
    \label{fig:SBGLS}
\end{figure}

 We also use the Stacked Bayesian Generalized Lomb-Scargle (SBGLS) periodogram \citep{2015A&A...573A.101M,2017A&A...601A.110M} in order to check the development of the signals with increasing number of data points (Fig.\,\,\ref{fig:SBGLS}). The three planetary signals are indeed stable in period and their probability is increasing with the number of data points as expected for coherent signals.

\section{Photometry}
\label{app:photofacilities}


 The used photometric facilities are summarized in Table~\ref{tab:phot-facilities}. They are:

ASH2. The ASH2 (Astrograph for South Hemisphere II) telescope) is a robotic 40~cm telescope with a CCD camera STL11000 2.7k$\times$4k, FOV 54$\times$82 arcmin$^2$. The telescope is at SPACEOBS\footnote{http://www.spaceobs.com/en} (San Pedro de Atacama Celestial Explorations Observatory), at 2450 m above the sea, 
located in the northern Atacama Desert in Chile. During the present work, only subframes with 40\% of the total FOV were used, resulting in a useful FOV of 21.6$\times$32.8 arcmin$^2$. Data are obtained as differential magnitude relative to a main reference star. A set of five check stars were selected and used for checking purposes. In total, 32 epochs in V and R bands were obtained in the period between July and October, 2018, with a time span of 97 days.

MONET-S. The 1.2 m MONET-South\footnote{https://monet.uni-goettingen.de/} telescope (MOnitoring NEtwork of Telescopes) is located at the SAAO (South African Astronomical Observatory), South Africa. It is equipped with a Finger Lakes ProLine 2k$\times$2k e2v CCD, FOV 12.6$\times$12.6 arcmin$^2$. 
We performed aperture photometry with the AstroImageJ package using a set of about five comparison stars. 43 epochs in the R band were collected during the Red Dots campaign.

AAVSO. Within the AAVSO collaboration, the Remote Observatory Atacama Desert (ROAD) \citep{2012SASS...31...75H} was used. This is also located at SPACEOBS. It is equipped with a 40~cm f/6.8 Optimized Dall-Kirkham (ODK) reflector from Orion Optics, UK, and a CCD camera  ML16803 4k$\times$4k from FLI, USA, equipped with Astrodon UBVRI 
photometric filters. Data were reduced using the LesvePhotometry package with two comparison stars, one for reference and another one for checking purposes. These observations were also collected simultaneously with the Red Dots campaign.

TESS. At the end of Red Dots campaign, GJ~1061 was also observed by the TESS\footnote{http://heasarc.gsfc.nasa.gov/docs/tess/} satellite \citep{Ricker2015JATIS...1a4003R}. This was carried out during two consecutive sectors of 27 days each, with a effective total time span of about 53 days between September and November, 2018, with the main aim of investigating possible transit signals.

MEarth-S. The MEarth project\footnote{https://www.cfa.harvard.edu/MEarth/} consists of two robotically controlled observatories dedicated to monitoring 
thousands of M-dwarf stars \citep{Berta2012AJ....144..145B}: MEarth-N (North), at Fred Lawrence Whipple Observatory (FLWO), USA, since 2008 and MEarth-S (South), at Cerro Tololo Inter-American Observatory (CTIO), Chile, since 2014.  Each array consists of eight identical 40~cm robotic telescopes (f/9 Ritchey-Chr{\'e}tien Cassegrain), each equipped with a 2k$\times$2k CCD camera, FOV 26$\times$26 arcmin$^2$, sensitive to red optical and near-infrared light. MEarth project generally uses an RG715 long-pass filter, except for the 2010-2011 season when an I$_{715-895}$ interference filter was choosen. In the case of GJ~1061, two time series are available from the MEarth-S project, which have been collected with telescopes number 11 (T11) (year 2017) and 13 (T13) (2016-2017).

ASAS-SN. The ASAS-SN project\footnote{http://www.astronomy.ohio-state.edu/asassn/} (All-Sky Automated Survey for Supernovae) \citep{Kochanek2017PASP..129j4502K} currently consists of 24 telescopes distributed on 6 units around the Globe, located in Hawaii (Haleakala Observatory), Chile (CTIO, 2 units), Texas (McDonald Observatory), South Africa (SAAO) and China. Each station consists of four 14~cm aperture Nikon telephoto lenses, each with a thermo-electrically cooled, back-illuminated, 2k$\times$2k, Finger Lakes Instruments, ProLine CCD camera. The field of view of each camera is roughly 4.5 deg$^2$ with a pixel scale of 8.0 arcsec.

\begin{table*}
    \caption{\label{tab:phot-facilities}Photometric observing facilities.}
    \centering
    \begin{tabular}{llccccl}
        \hline
        \hline
Acronym     &   Location                   &    Tel        &     FOV                   &   CCD           &         Scale           &  Band(s) \\
            &                              &    (m)        & (arcmin$^2$)              &                 &   (arcsec pix${^-1}$)   &          \\
        \hline
ASH2       & SPACEOBS, Chile               & 0.40          & 54.0$\times$82.0                                & 2.7k$\times$4k   &   1.20   &  V,R   \\
MONET-S    & SAAO, South Africa            & 1.20          & 12.6$\times$12.6                                & 2k$\times$2k     &   0.37   &  R     \\
AAVSO      & SPACEOBS, Chile               & 0.40          & 47$\times$47                                    & 4k$\times$4k     &   0.69  &  V     \\
TESS       & TESS satellite                & 4$\times$0.105 & 4$\times$(24$^{\circ}\times$24$^{\circ}$) &  4k$\times$4k    &  21.1    &  TESS  \\
MEarth-S   & CTIO, Chile                   & 0.40          & 26.0$\times$26.0                                & 2k$\times$2k     &   0.76   &  RG715 \\
ASAS-SN    & worldwide                     & 4$\times$0.14  & 4.5 deg$^2$                                     & 2k$\times$2k     &   8.0    &  V     \\
        \hline
    \end{tabular}
\end{table*}
\begin{table}
    \centering
    \caption{Ground-based photometry, full table in online data.}
    \label{tab:GB_phot}
    \begin{tabular}{crr}
    \hline
    \hline
    BJD & rel. mag & $\sigma$ rel. mag \\
    \hline
    \multicolumn{3}{c}{ASH2 R}\\
    \hline
2458309.830903 &	0.0062	& 0.0016 \\
2458310.834277 &	0.0048	& 0.0024 \\
    \hline
    \multicolumn{3}{c}{ASH2 V}\\
    \hline
2458309.829251 &	0.0001 &	0.0020 \\
2458310.835848 &	0.0028 &	0.0040 \\
    \hline
    \multicolumn{3}{c}{MONET-S R}\\
    \hline
2458308.622806 &	0.0022 &	0.0010 \\
2458318.563727 &	0.0160 &	0.0011 \\
    \hline
    \multicolumn{3}{c}{AAVSO R}\\
    \hline
2458308.901969 &	0.0119 &	0.0017 \\
2458309.899256 &	0.0137 &	0.0015 \\
    \hline
    \multicolumn{3}{c}{MEarth RG715}\\
    \hline
2457905.943122 &	-0.0043 &	0.0029 \\
2457906.939185 &	-0.0067 &	0.0073 \\
\hline
    \end{tabular}
\end{table}

\begin{figure}
    \centering
    \includegraphics[width=1\linewidth]{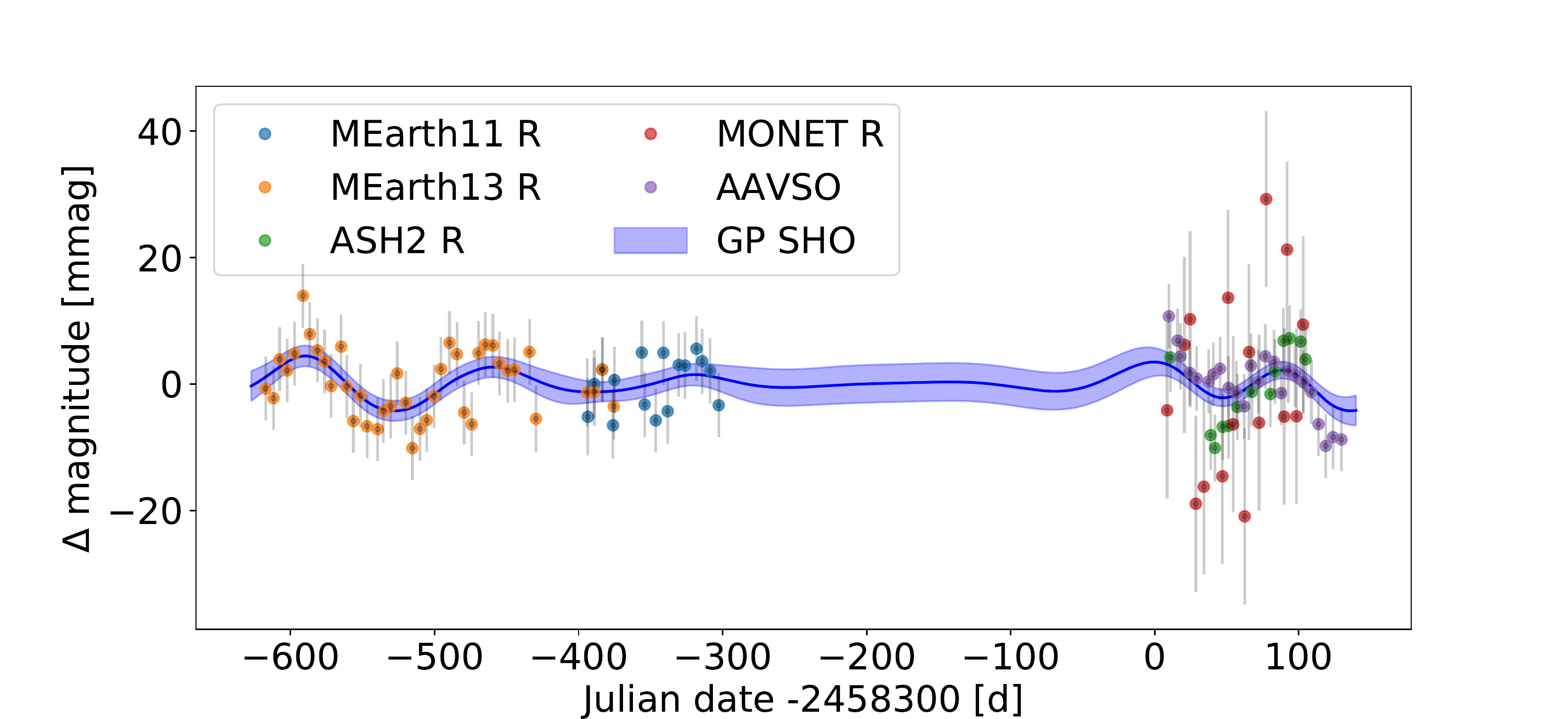}
    \caption{Photometric data, shown in five day bins together with a stochastically driven, damped harmonic oscillator model using the {\em SHO} kernel with 130\,d period and 20\,d damping time.}
    \label{fig:photo}
\end{figure}
\begin{figure}
    \centering
    \includegraphics[width=1\linewidth]{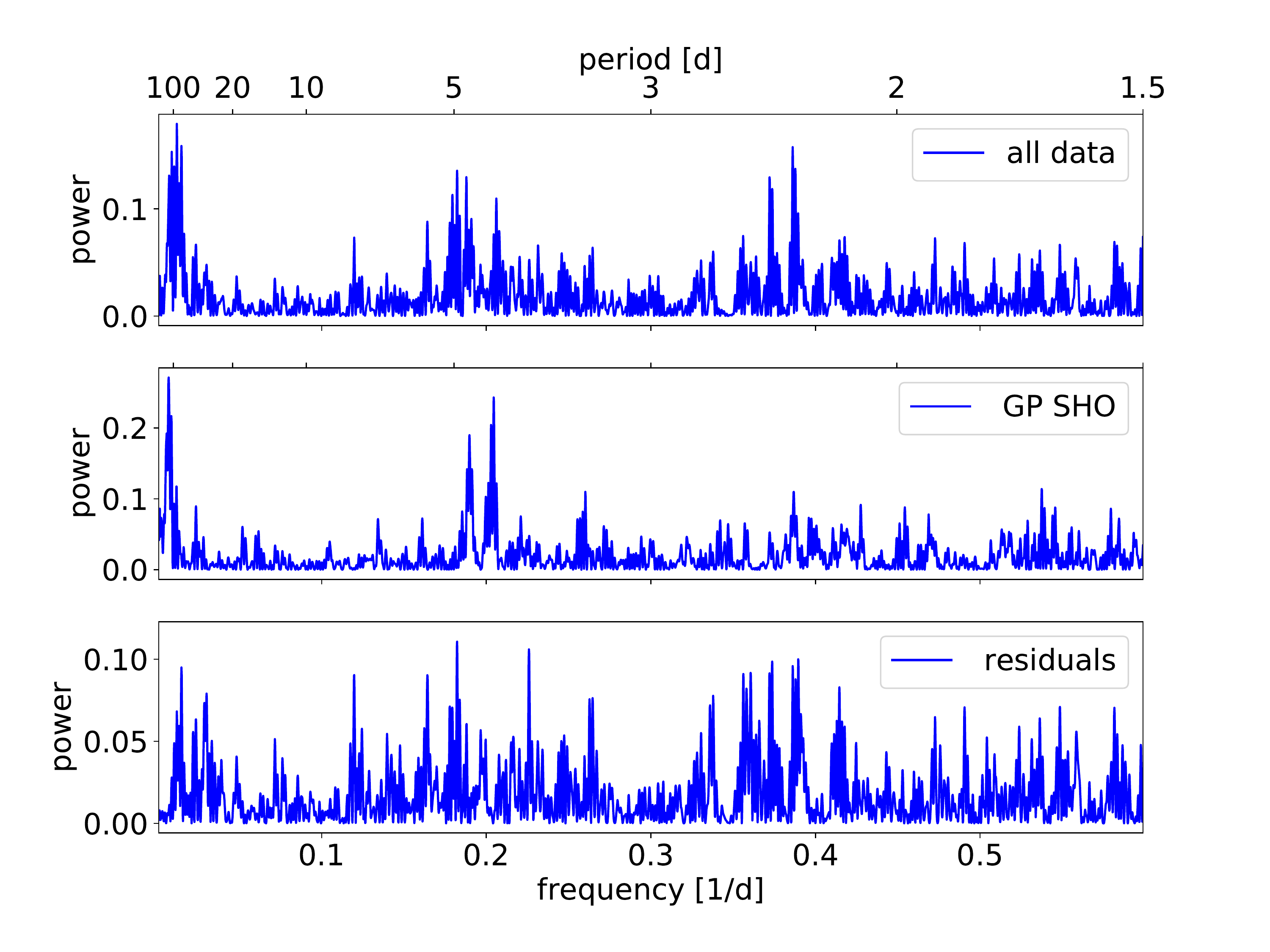}
    \caption{Periodogram of photometric data (top), of the {\em SHO} kernel (middle) and of the residuals (bottom).}
    \label{fig:photoperiodogram}
\end{figure}

The combined photometric data are listed in Table\,\ref{tab:GB_phot} and shown in Fig.\,\,\ref{fig:photo} together with a model using the {\em SHO}-kernel from {\em celerite}. The signal at about 130\,d is well described with this quasi-periodic oscillations as visible from the periodograms in Fig.\,\,\ref{fig:photoperiodogram}. The analysis is presented in the main text in Sect.\,\ref{sec:Photo}.

\section{Activity indicators}
\begin{table}
    \centering
    \caption{Equivalent widths obtained by {\em TERRA} in the HARPS post-upgrade, data. The flag $0$ indicates rejected data due to possible flares. Full table in online data.}
    \label{tab:EWs}
    \begin{tabular}{crrrr}
\hline
\hline
BJD & H$\alpha$ & Na D1 & Na D2 & flag \\
    & \AA & \AA & \AA & \\
\hline
2458020.814834 & -0.12 & 1.17 & 0.95 & 1 \\
2458040.854023 & -0.16 & 1.17 & 0.90 & 1 \\
2458043.783336 & -0.02 & 1.23 & 0.97 & 1 \\
2458052.765154 & -0.03 & 1.18 & 0.93 & 1 \\
       \hline
    \end{tabular}
\end{table}

\begin{figure}
    \centering
    \includegraphics[width=1\linewidth]{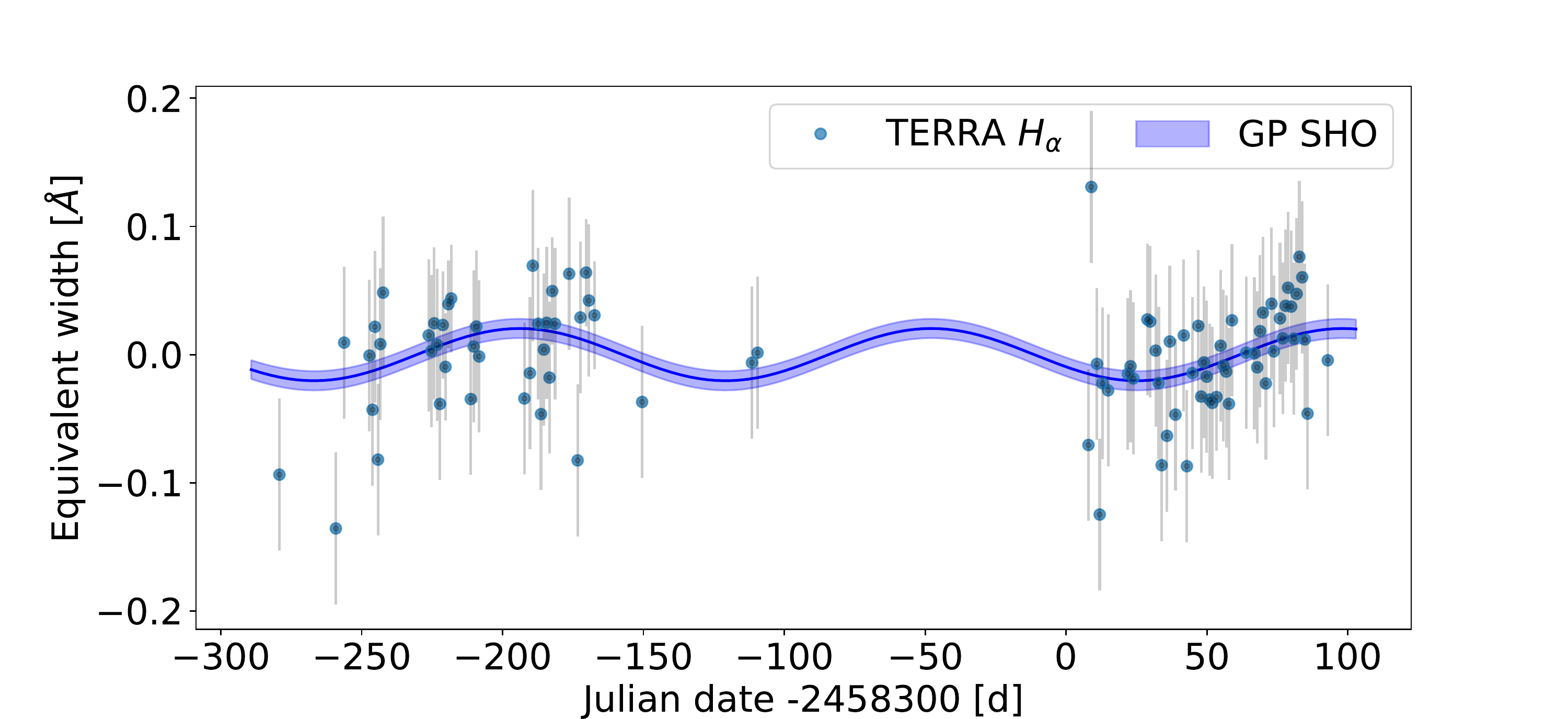}
    \caption{H$\alpha$ data, shown with a stochastically driven, damped harmonic oscillator model using the {\em SHO} kernel with 130\,d period and 20\,d damping time.}
    \label{fig:Halpha}
\end{figure}

\begin{figure}
    \centering
    \includegraphics[width=1\linewidth]{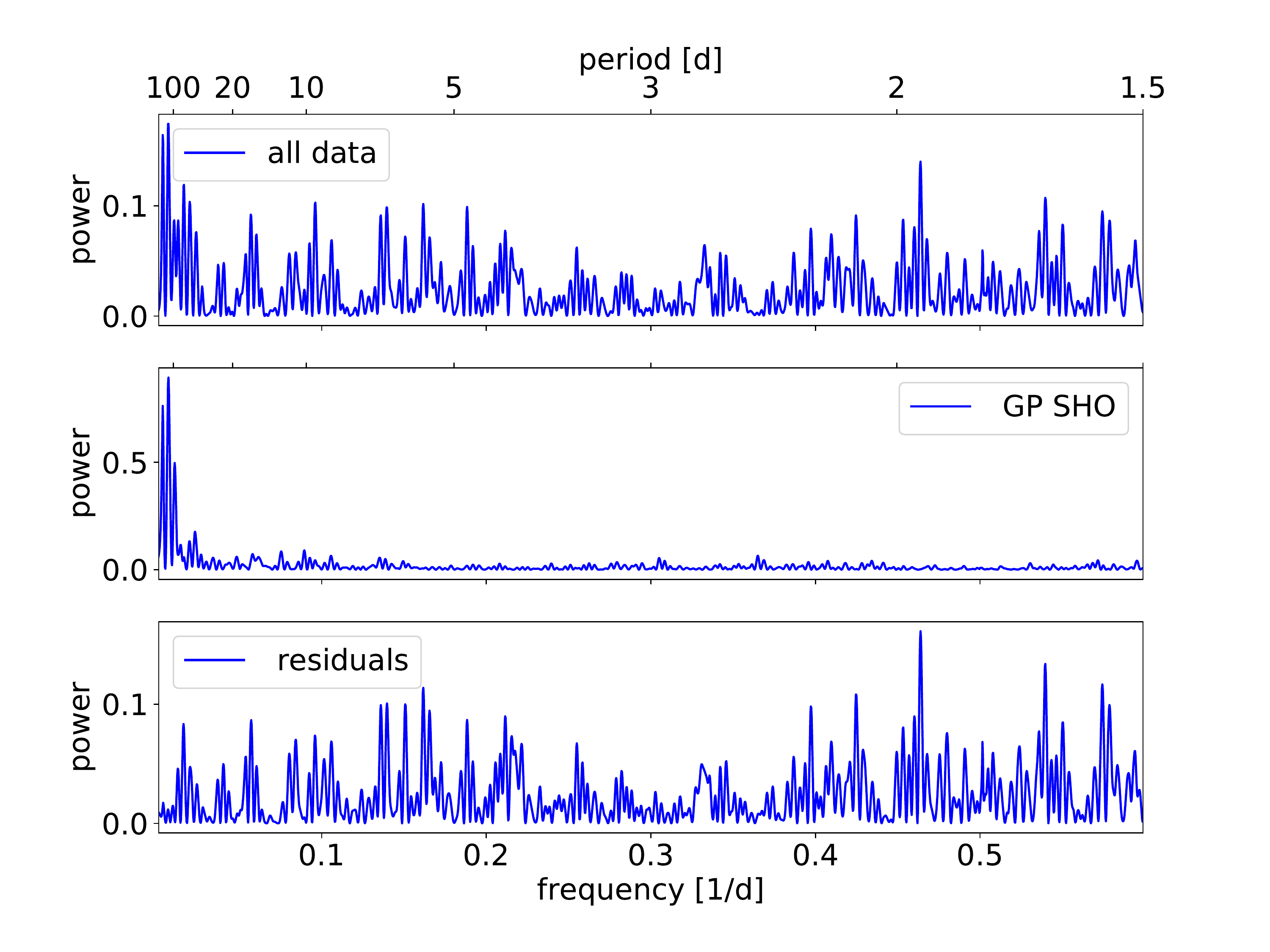}
    \caption{Periodogram of H$\alpha$ data (top), of the {\em SHO} kernel (middle) and of the residuals (bottom).}
    \label{fig:Halphaperiodogram}
\end{figure}

\begin{figure}
    \centering
    \includegraphics[width=1\linewidth]{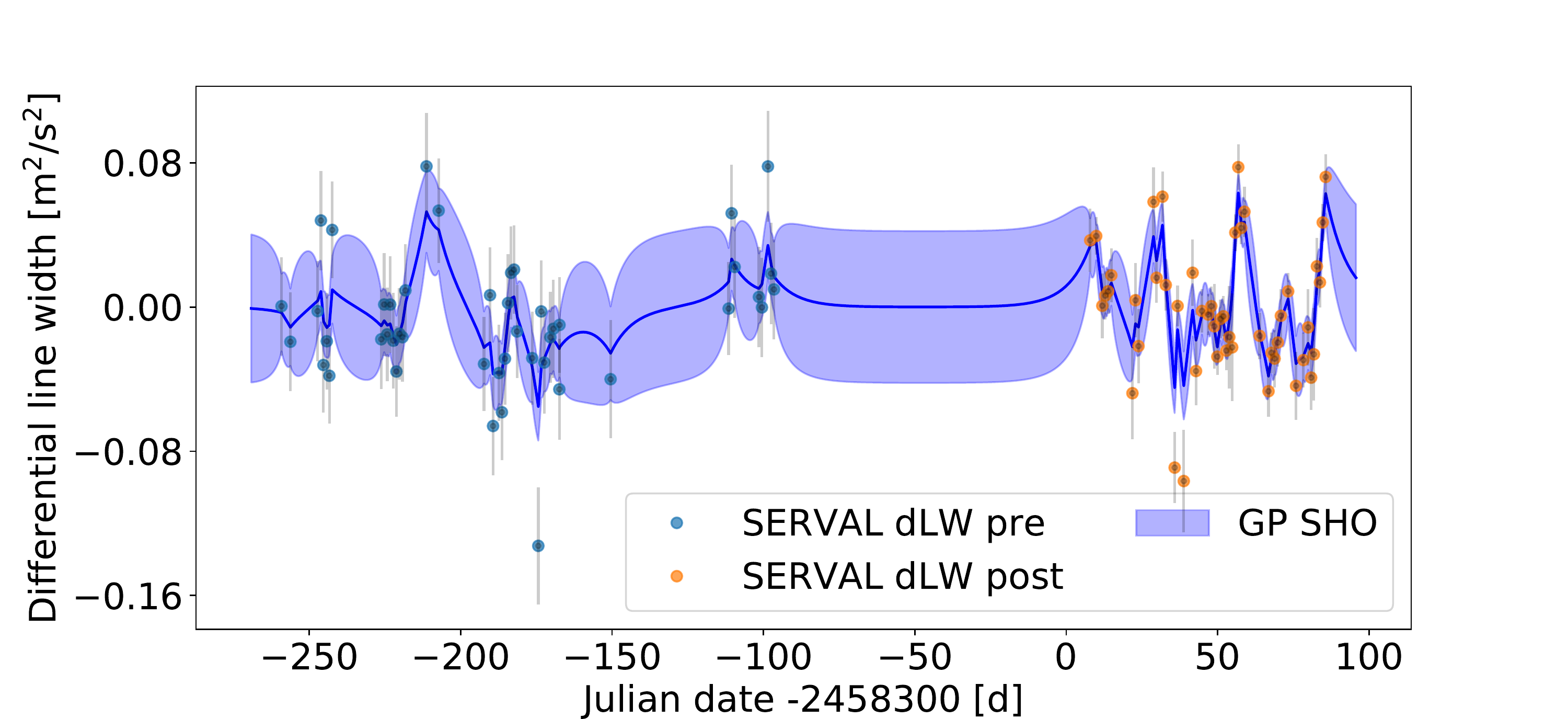}
    \caption{Differential line width data, shown together using the {\em REAL} kernel with 3.5\,d decay time.}
    \label{fig:dlw}
\end{figure}

\begin{figure}
    \centering
    \includegraphics[width=1\linewidth]{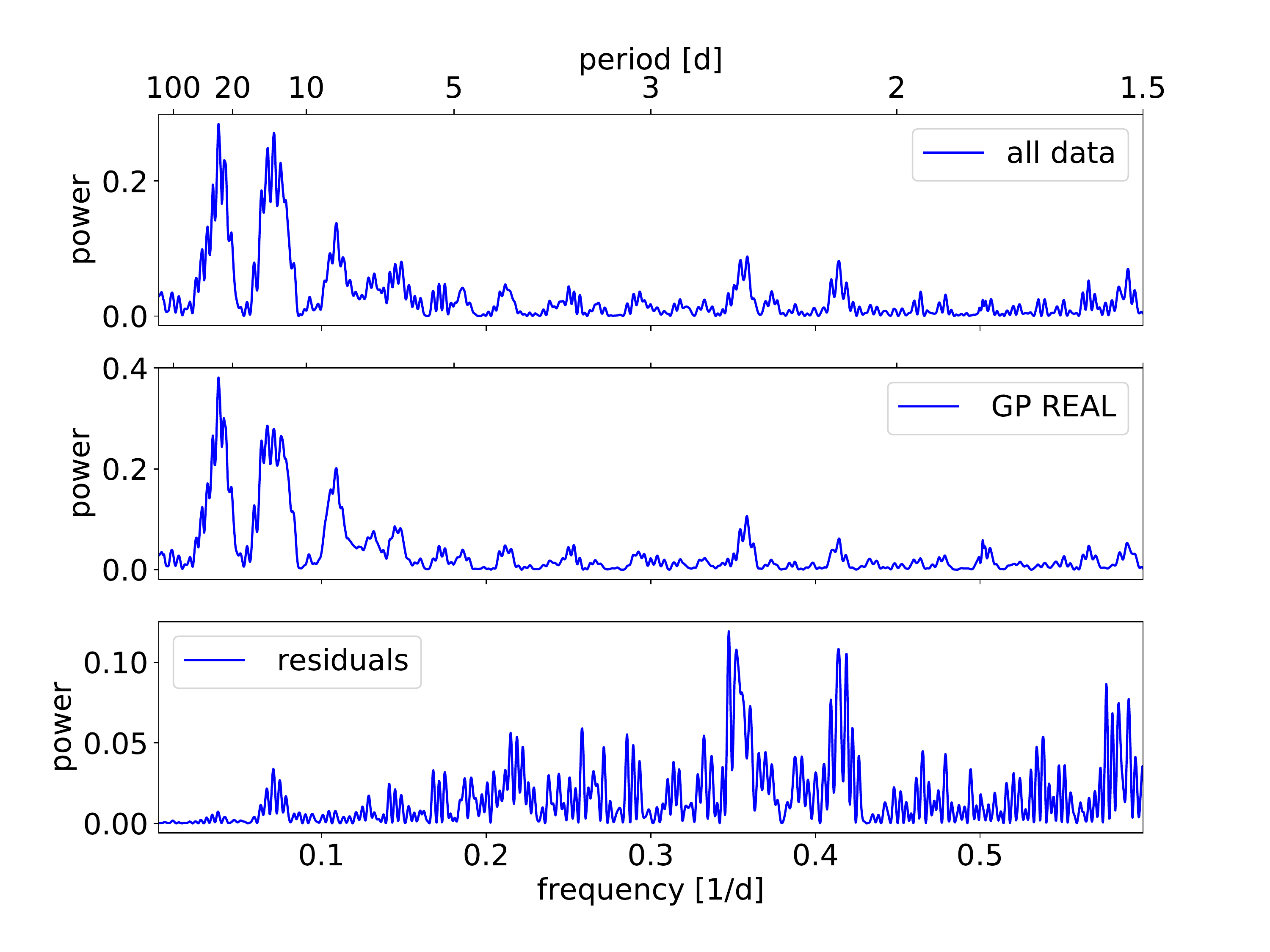}
    \caption{Periodogram of differential line width data (top), of the {\em REAL} kernel (middle) and of the residuals (bottom).}
    \label{fig:dlwperiodogram}
\end{figure}

{\em TERRA} and {\em SERVAL} provide various spectroscopic activity indicators, among those the  H$\alpha$ as well as NaD equivalent width and the differential line width. These measurements are listed in Tables\,\ref{tab:EWs} and \ref{tab:RV Hpost}. The spectroscopic activity indicators have been investigated for signals with periods at the potential planetary periods to avoid false positive planet detections as well as for the detection of the stellar rotation. In Fig.\,\,\ref{fig:dlw} we show the fit of the differential line width, in Fig.\,\,\ref{fig:Halpha} for the H$_\alpha$ equivalent width. The periodograms of the data, signals and residuals are shown in Figs.\,\ref{fig:dlwperiodogram} and \ref{fig:Halphaperiodogram}. The analysis is presented in the main text in Sect.\,\ref{sect:Activity}.



\bsp	
\label{lastpage}
\end{document}